\def\ts     {\thinspace}
\def\kms    {\ifmmode{{\rm \ts km\ts s}^{-1}}\else{\ts km\ts s$^{-1}$}\fi}
\def\msol   {\ifmmode{{\rm M}_{\odot} }\else{M$_{\odot}$}\fi}
\def\lsol   {\ifmmode{{\rm L}_{\odot}}\else{${\rm L}_{\odot}$}\fi}
\def\Lsol   {\ifmmode{{\rm L}_{\odot}}\else{${\rm L}_{\odot}$}\fi}
\def\lfir   {\ifmmode{{\rm L}_{\rm FIR}}\else{${\rm L}_{\rm FIR}$}\fi}
\def\Lfir   {\ifmmode{{\rm L}_{\rm FIR}}\else{${\rm L}_{\rm FIR}$}\fi}
\def\zsol   {\ifmmode{{\rm Z}_{\odot}}\else{Z$_{\odot}$}\fi}
\def\etal   {{\rm et\ts al.\ }}
\def\aco    {\ifmmode{{\rm CO}(J\!=\!1\! \to \!0)}\else{{\rm CO}($J$=1$\to$0)}\fi}
\def\bco    {\ifmmode{{\rm CO}(J\!=\!2\! \to \!1)}\else{{\rm CO}($J$=2$\to$1)}\fi}
\def\cco    {\ifmmode{{\rm CO}(J\!=\!3\! \to \!2)}\else{{\rm CO}($J$=3$\to$2)}\fi}
\def\dco    {\ifmmode{{\rm CO}(J\!=\!4\! \to \!3)}\else{{\rm CO}($J$=4$\to$3)}\fi}
\def\eco    {\ifmmode{{\rm CO}(J\!=\!5\! \to \!4)}\else{{\rm CO}($J$=5$\to$4)}\fi}
\def\fco    {\ifmmode{{\rm CO}(J\!=\!6\! \to \!5)}\else{{\rm CO}($J$=6$\to$5)}\fi}
\def\gco    {\ifmmode{{\rm CO}(J\!=\!7\! \to \!6)}\else{{\rm CO}($J$=7$\to$6)}\fi}
\def\hco    {\ifmmode{{\rm CO}(J\!=\!8\! \to \!7)}\else{{\rm CO}($J$=8$\to$7)}\fi}
\def\ico    {\ifmmode{{\rm CO}(J\!=\!9\! \to \!8)}\else{{\rm CO}($J$=9$\to$8)}\fi}
\def\jco    {\ifmmode{{\rm CO}(J\!=\!10\! \to \!9)}\else{{\rm CO}($J$=10$\to$9)}\fi}
\def\kco    {\ifmmode{{\rm CO}(J\!=\!11\! \to \!10)}\else{{\rm CO}($J$=11$\to$10)}\fi}
\def\ci     {\ifmmode{{\rm C}{\rm \small I}}\else{C\ts {\scriptsize I}}\fi}
\def\hi     {\ifmmode{{\rm H}{\rm \small I}}\else{H\ts {\scriptsize I}}\fi}
\def\hh     {\ifmmode{{\rm H}_2}\else{H$_2$}\fi}
\def\cone {\ifmmode{{\rm C}{\rm \small I}(^3\!P_1\!\to^3\!P_0)}
     \else{C\ts {\scriptsize I}{\small$(^3\!P_1\!\to^3\!\!\!P_0)$}}\fi}
\def\ctwo {\ifmmode{{\rm C}{\rm \small I}(^3\!P_2\!\to^3\!P_1)}
     \else{C\ts {\scriptsize I}{\small$(^3\!P_2\!\to^3\!\!\!P_1)$}}\fi}
\def\cij {\ifmmode{{\rm C}{\rm \small I}\,(^3P_i\to^3P_j)}
\else{C\ts {\scriptsize I}\,{\small$(^3P_i\to^3P_j)$}}\fi}
\def\cii    {\ifmmode{{\rm C}{\rm \small II}}\else{C\ts {\scriptsize II}}\fi}
\def\tex {\ifmmode{{T}_{\rm ex}}\else{$T_{\rm ex}$}\fi}
\def\tmb {\ifmmode{{T}_{\rm mb}}\else{$T_{\rm mb}$}\fi}
\def\tkin {\ifmmode{{T}_{\rm kin}}\else{$T_{\rm kin}$}\fi}
\def\microns {\ifmmode{\mu{\rm m}}\else{$\mu$m}\fi}
\def\nhh   {\ifmmode{n({\rm H}_2)}\else{$n$(H$_2$)}\fi}
\def\gradv {\ifmmode{(dv/dr)}\else{$(dv/dr)$}\fi}
\def\CO10{{\hbox {CO(1--0)}}}

\def\,{\thinspace}
\def\Msun{M$_\odot$}
\def\msun{M$_\odot$}
\def \Kkmspc{K\,\kms\,pc$^2$} 
\def \kkmspc{K\,\kms\,pc$^2$} 
\documentclass[twocolumn]{aa}    
\usepackage{graphicx}
\begin{document}
 \titlerunning{Highly-excited CO in APM\,08279+5255 at $z=3.9$}
 \title{Highly-excited CO emission in APM\,08279+5255 at z=3.9}

   \author{A. Wei\ss
          \inst{1}
          \and
          D. Downes
          \inst{2}
          \and
          R. Neri
          \inst{2}
	  \and
          F. Walter
          \inst{3}
          \and
          C. Henkel
          \inst{1}
	  \and
	  D.J. Wilner
          \inst{4}
          \and
	  J. Wagg
	  \inst{4,5}
	  \and
	  T. Wiklind
	  \inst{6}
          }

   \institute{MPIfR, Auf dem H\"ugel 69, 53121 Bonn, Germany
         \and
             IRAM, Domaine Universitaire, 38406 St-Martin-d'H\`eres, France
         \and
             MPIA, K\"onigstuhl 17, 69117 Heidelberg, Germany
	 \and
	     Harvard-Smithsonian Center for Astrophysics, Cambridge, MA, 02138
	 \and
	     Instituto Nacional de Astrofisica, \'Optica y Electr\'onica
	     (INAOE), Aptdo.\ Postal 51 y 216, Puebla, Mexico
         \and
	     ESA-Space Telescope Division, STScI, 3700 San Martin Drive,
	     Baltimore, MD 21218, USA
             }

   \date{}
   \abstract{
We report the detection of the CO
4--3, 6--5, 9--8, 10--9, and 11--10 lines in the Broad Absorption Line
quasar APM\,08279+5255 at $z=3.9$ using the IRAM 30\,m telescope.  We
also present IRAM PdBI high spatial resolution observations of the CO 4--3 and 9--8
lines, and of the 1.4\,mm dust radiation as well
as an improved spectrum of the HCN(5--4) line. Unlike CO in other QSO
host galaxies, the CO line SED of APM\,08279+5255 rises up to the CO(10--9)
transition.  The line fluxes in the CO ladder and the dust continuum fluxes are best
fit by a two component model, a ``cold'' component at $\sim 65$\,K
with a high density of $n$(H$_2$)= $1\cdot 10^5$\,cm$^{-3}$, and a ``warm'', $\sim 220$\,K component with a density of
$1\cdot 10^4$\,cm$^{-3}$. We show that IR pumping via the 14\microns\ bending mode of HCN is the
most likely channel for the HCN excitation. From our models we find, that the CO(1--0)
emission is dominated by the {\it dense} gas component which implies
that the CO conversion factor is higher than usually assumed for high-z galaxies with
$\alpha\,\approx\,5\,\msol\ ({\rm K\,\kms\,pc^2})^{-1}$.  Using
brightness temperature arguments, the results from our high-resolution
mapping, and lens models from the literature, we argue that the
molecular lines and the dust continuum emission arise from a very
compact ($r\approx100-300$\,pc), highly gravitationally magnified
($m= 60-110$) region surrounding the central AGN.  Part of the
difference relative to other high-$z$ QSOs may therefore be due to the
configuration of the gravitational lens, which gives us a
high-magnification zoom right into the central 200-pc radius of
APM\,08279+5255 where IR pumping plays a significant role for the excitation 
of the molecular lines.

}
   \maketitle

\section{Introduction} 

The existence of massive reservoirs of molecular gas at high redshifts has
now been established in quasars, submillimeter galaxies and radio galaxies
out to the highest redshifts (see review by Solomon \& Vanden Bout \cite{solomon05}
and references therein). It is now of interest to study in detail the
excitation conditions of the molecular gas in these high-redshift
systems to search for differences of the gas properties among high-z sources and 
to relate them to the properties of their host galaxies.
One way to do this is by studying multiple CO transitions from individual
key sources (hereafter referred to as `CO line spectral energy
distributions (SEDs)'). We have recently reported observations for the
gravitationally lensed galaxy SMM J16359+6612 (Weiss \etal
\cite{weiss05a}).  Here we focus on new observations of one of the
brightest quasars at high redshifts, the broad absorption line (BAL) 
quasar APM\,08279+5255 at $z=3.9$ (Irwin \etal \cite{irwin98}).

\begin{figure*}[ht]
\centering
\includegraphics[angle=0,width=16.0cm]{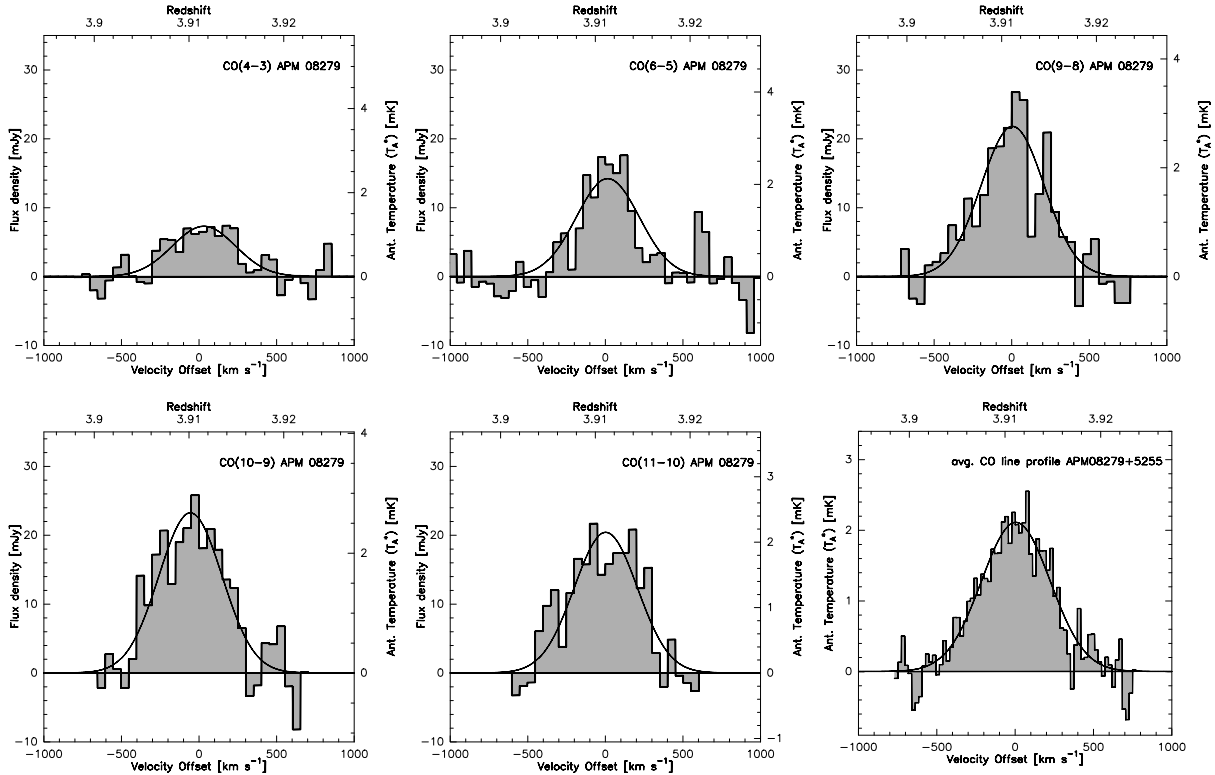}
\caption[30m Spectra]{
IRAM 30\,m telescope spectra of CO 4--3, 6--5, 9--8, 10--9, 11--10 and the average CO
spectrum from APM\,08279+5255, with Gaussian fit profiles superimposed.  Velocity scales are
relative to a redshift of $z=3.911$. The velocity resolution for individual
spectra is 50\,\kms.  For the average spectrum, the velocity resolution is 22\,\kms.  The individual spectra 
are plotted on the same scale in flux density ({\it left axis}).
}
\label{pico-spectra} 
\end{figure*}

\noindent APM\,08279+5255 has a stunning apparent luminosity of $L_{\rm bol}=
7\times 10^{15}$\,\lsol , one of the highest in the universe (Irwin
\etal \cite{irwin98}).  High-resolution imaging and spectroscopy by
Ledoux \etal (\cite{ledoux98}) initially showed the optical source to
have two images, $A$ and $B$, separated by $0^{\prime\prime}\!.38$,
and showing the same spectrum, indicating that APM\,08279 is
gravitationally lensed.  Subsequently Ibata \etal (\cite{ibata99}) and
Egami \etal (\cite{egami00}) found a third image, $C$,
$0^{\prime\prime}\!.1$ north of $A$, implying that this is a rare
cusped lens with an odd number of images. The multiple image structure
has also been seen in X-rays (Chartas \etal \cite{chartas02}).  The
lens modeling by Egami \etal (\cite{egami00}) indicates a high
magnification factor of $\sim$100 at optical wavelengths.  The
object's true bolometric luminosity would therefore be $\sim 7\times
10^{13}$\,\Lsol , still making it one of the most luminous quasars.
APM\,08279 was also detected in the mm and submm dust continuum, and
its SED peaks near $\sim30$\,\microns\ (restframe), consistent with a single component dust fit with $T_{\rm dust} =
220$\,K (Lewis \etal \cite{lewis98}) --- significantly warmer than the dust emission in
other high-$z$ QSOs with typical temperatures of 50 -- 80\,K (see
e.g. Beelen \etal \cite{beelen06}).

\noindent Molecular gas in APM\,08279 object has been first detected through
observations of the CO(4--3) and CO(9--8) lines (Downes \etal 1999;
hereafter D99) using the IRAM interferometer.  These high-$J$ CO
detections were followed up by observations of the CO(1--0) and CO(2--1)
lines at the VLA (Papadopoulos \etal \cite{papado01}; Lewis \etal
\cite{lewis02a}).  Based on the earlier observations, initial
gravitational lens models suggested the CO lines arise in a compact
($r\approx 400$\,pc) disk surrounding the quasar and that the CO was
magnified by a factor of 5 to 20 (D99; Lewis \etal \cite{lewis02a},b).
Based on the morphology of the low--J CO transitions Papadopoulos \etal (\cite{papado01})
suggested the presence of extended molecular gas emission in this
source. Subsequent GBT observations by Riechers \etal \cite{riechers06} and sensitive
new high--resolution VLA imaging (Riechers \etal in prep), however, do not show
evidence for such a resolved molecular gas component: the CO emission appears
to be co--spatial to the emission seen in the NIR. 

\noindent While the size and mass ($10^9$ to $10^{10}$\,\msol) of the molecular gas
toroid in APM\,08279 is similar to those derived for other high--z QSOs
(Solomon \& Vanden Bout 2004) the mere detection of the CO(9--8) line in
APM\,08279 already indicates that the physical properties of its molecular
gas phase are more extreme than what is found in other high-$z$ QSOs.  
Another sign of these more extreme conditions is the detection of
HCN(5--4) in this source, which also indicates high gas densities (Wagg
\etal \cite{wagg05}).  To better constrain the excitation of the molecular
gas, we here present a detailed study of the CO line SED in this source.
The observations were obtained at the IRAM 30\,m telescope (CO 4--3, 6--5,
9-8, 10--9, and 11--10 lines).  We also present Plateau de Bure
interferometer long-baseline observations of the CO 4--3, 9--8, and
1.4\,mm dust radiation that have higher resolution than the earlier Bure
maps of this source (D99).

\section{Observations and Results}

\subsection{Pico Veleta}
We observed APM\,08279+5255 with the IRAM 30\,m telescope on Pico
Veleta, Spain, in the winters of 2004/05 and 2005/06, in
excellent weather. We used the AB and CD receiver setups, with the A/B
receivers tuned to CO(4--3) (3\,mm band) and either CO(9--8) or
CO(10--9) (lower 1\,mm band) and the C/D receivers tuned to CO(6--5)
(2\,mm band) and CO(11--10) (upper 1\,mm band).  System temperatures
were typically $\approx$\,140\,K, 180\,K, and 320\,K ($T_a^*$) for the 3, 2, and
1\,mm observations.  We observed with a wobbler rate of 0.5\,Hz and
a wobbler throw of $50''$ in azimuth, and frequently checked the pointing, which
was stable to within $3''$ in all runs.  We calibrated every 12\,min with standard
hot/cold load absorbers, and estimate fluxes to be accurate to
$\pm$10\% at 3 and 2\,mm and $20\%$ at 1\,mm.  Spectrometers were the
$512\times 1$\,MHz filter banks for the 3\,mm receivers, and the
$256\times 4$\,MHz filter banks for the 1.3\,mm receivers.  In the
data processing, we dropped all scans with distorted baselines,
subtracted linear baselines from the remaining spectra, and then
rebinned to a velocity resolution of 50\,\kms . 
Table\,\ref{linepara} summarizes the observing parameters, and 
Fig.\,\ref{pico-spectra}  shows the spectra.  

\noindent The most remarkable
result of our CO line SED study of APM\,08279 at the 30\,m telescope
is that we detect the CO(10--9) and the CO(11-10) lines 
(Fig.~\ref{pico-spectra}). To search for differences in the line
profile between the mid-$J$ and high-$J$ lines of CO we averaged the
CO(9--8) CO(10--9) and CO(11--10) spectra with equal weight for each
line. The linewidth from this high-$J$ CO spectrum is $500\pm20$\,\kms
, which agrees with the value derived from CO(4--3) alone
($480\pm35$\,\kms ; D99). From the average profile of all CO lines
(Fig.~\ref{pico-spectra}),  we derive a CO linewidth of
$500\pm26$\,\kms\ and a CO redshift of $z=3.9112\pm0.0004$. Our new
30\,m CO(4--3) line intensity agrees with both the earlier (D99) and
our new PdBI measurement. The 30\,m CO(9--8) line intensity also
agrees well with our new PdBI observations.  Both new measurements
yield a flux density slightly higher, but consistent, within the
calibration error, with the previous measurements (D99).

\subsection{Plateau de Bure}
\subsubsection{CO observations}

We also observed the CO(4--3) and CO(9--8) lines with the IRAM
Plateau de Bure interferometer in the long--baseline $A$ and $B$ configurations. 
The resulting data have an equivalent 6-antenna on-source integration
time of 15\,h, with baselines from 24 to 410\,m,
giving naturally-weighted synthesized beams of $1^{\prime\prime}\!.7 \times 1^{\prime\prime}\!.6$ at
3.2\,mm and $0^{\prime\prime}\!.85 \times 0^{\prime\prime}\!.71$ at
1.4\,mm. Receiver temperatures were 45 to 65\,K at both wavelengths. The spectral correlators covered
910\,\kms\ at 3.2\,mm and 760\,\kms\ at 1.4\,mm, with resolutions of
8.0 and 3.6\,\kms\ respectively. Amplitudes were calibrated with 3C84,
3C454.3, 3C273, and MWC349, and phases were calibrated with
IAP~0749+540 and IAP~0804$+$499. 

\begin{figure}[ht] 
\centering
\includegraphics[angle=0,width=8.0cm]{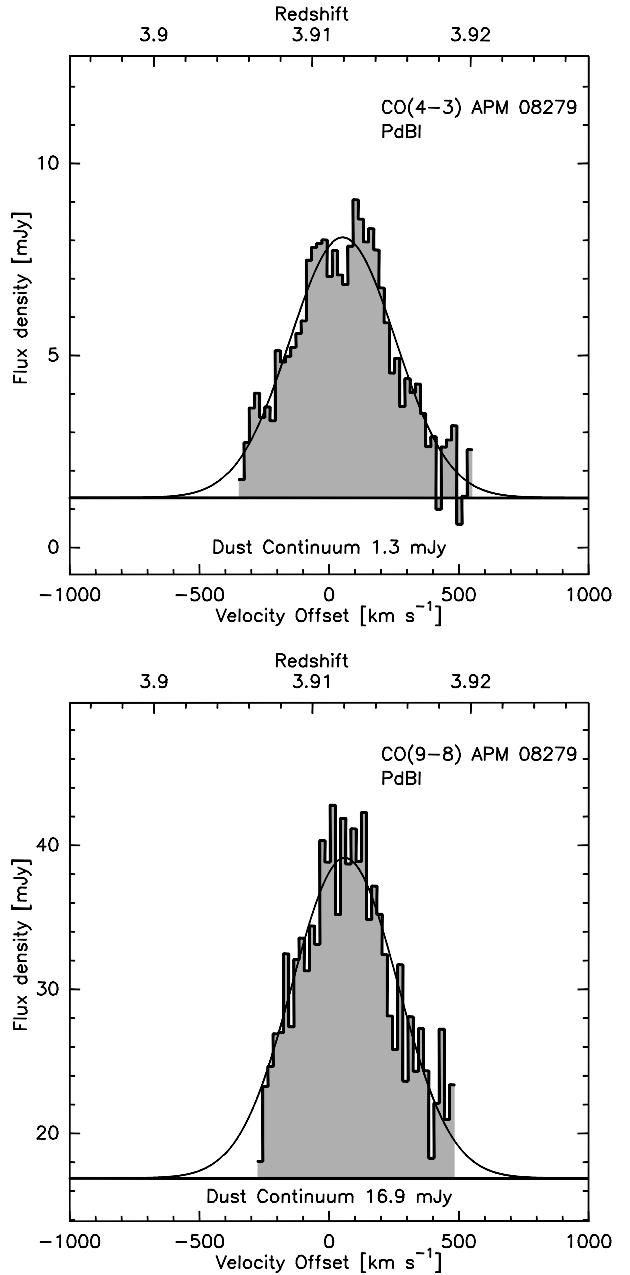}
\caption[PdB CO43 and CO98 Spectra]
{Integrated CO(4--3) and CO(9--8) spectra from APM\,08279+5255 obtained with the
IRAM Interferometer. For both spectra the velocity offsets are given relative to a
redshift of $z=3.911$. {\it Upper panel:} CO(4--3) profile above the dust
continuum of 1.3\,mJy. The channel width
is 20\,\kms with an rms noise of 0.7\,mJy.
{\it Lower panel:} CO(9--8) profile above the dust continuum.  
The channel width is 20\,\kms with an rms noise of 1.7\,mJy.
The dust continuum is 16.9\,mJy. }
\label{pdblines} 
\end{figure}

\noindent To obtain high signal-to-noise spectra, we add the older
short-baseline configuration $D$ data from D99 to the long-baseline data.
These combined data have equivalent 6-antenna integration times of
26\,h at 3.2\,mm and 23\,h at 1.4\,mm and naturally-weighted
synthesized beams of $2^{\prime\prime}\!.9 \times 2^{\prime\prime}\!.4$ at
3.2\,mm and $1^{\prime\prime}\!.4 \times 1^{\prime\prime}\!.0$ at 1.4\,mm.
The integrated CO spectra from this data cube are shown in
Fig.\,\ref{pdblines}.  Such spectra of the total flux can be
obtained by either spatially integrating over the source map, or by
doing {\it u,v}-plane fits to the data in each channel, with the
source centroid and source size fixed to $0^{\prime\prime}\!.6$ (see below).
The fit then gives the zero-spacing flux (total flux) in each channel.
Both methods give the same result. For comparison with
the 30\,m spectra, we also fit these interferometer line profiles with
single Gaussians. Table\,\ref{linepara}
summarizes our results. The CO linewidth and redshift derived from the average of the two 
PdBI CO spectra give $475\pm17$\,\kms\ and $z=3.9119\pm0.0002$. This redshift is
higher than the corresponding value derived from the average 30\,m spectrum but in 
good agreement with the HCN redshift measured with the PdBI. We attribute 
these differences to the lower S/N ratio and remaining baseline instabilities in the 30\,m data.
The continuum fluxes at 3.2\,mm and 1.4\,mm
are 1.3\,mJy and 16.9\,mJy respectively - in agreement with the earlier
data (D99). 

\begin{figure} \centering
\includegraphics[angle=-90,width=8.0cm]{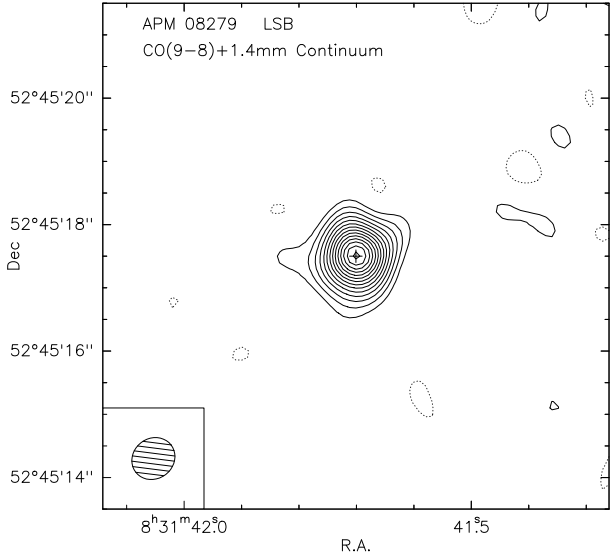}
\caption[CO98 Map]
{Map of an $8''$ field around APM~08279 made with the
IRAM Interferometer at 1.4\,mm.  The signal is the CO(9--8) line
integrated over 760\,\kms , plus the 1.4\,mm dust emission at the
highest spatial resolution (uniform
weighting,  beam: $0^{\prime\prime}\!.71
\times 0^{\prime\prime}\!.64$ ({\it lower left inset})).
Contours start at
3$\sigma= 1.35$\,mJy beam$^{-1}$ and increase in steps of 3$\sigma$.
The peak is 19.4\,mJy beam$^{-1}$ (43$\sigma$) and the spatially-integrated flux
is 33.9\,mJy. }
\label{1mmlsbmap} 
\end{figure}

\noindent To determine the astrometry and apparent size of the dust and CO 
emitting region we use the uniform weighted 1\,mm data which has a spatial
resolution of $0^{\prime\prime}\!.71 \times 0^{\prime\prime}\!.64$.
This CO(9--8) and 1.4\,mm dust maps (Fig.\,\ref{1mmlsbmap})
yield a source position (Table\,\ref{tablepos})  that agrees within the errors
with the earlier result (D99) and coincides within $0^{\prime\prime}\!.3$ with the
optical quasar (revised optical position from Irwin \etal \cite{irwin98}), and with
the non-thermal radio source detected at 1.4 GHz (Ivison \cite{ivison06}).

%
\begin{figure}[h] 
\centering
\includegraphics[angle=-90,width=8.0cm]{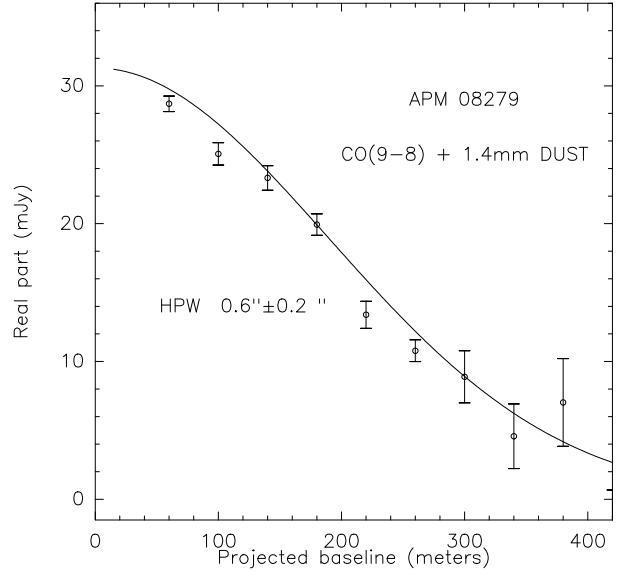}
\caption[Visibility-CO98+Dust]
{Size measurement with the IRAM Interferometer: visibility amplitudes
of the signal in the receivers' lower sideband at 1.4\,mm.  The
signal is the CO(9--8) line integrated over 760\,\kms , plus the
1.4\,mm dust emission.  The plot shows the real part of the visibility
amplitude vs.\ the projected antenna spacing, for $u,v$-plane data
averaged in circular bins 40\,m wide, with error bars of $\pm
1\sigma$.  The solid curve is a circular Gaussian fit 
with half-power width $0^{\prime\prime}\!.6 \pm 0^{\prime\prime}\!.2$. 
}
\label{visi-1mmlsb} 
\end{figure}

\noindent In its long-baseline $A$ and $B$ configurations, the IRAM
Interferometer starts to resolve the source as the peak flux is no
longer equal to the total flux. But the beam size is still too
large to show whether the mm-source is in three images, as in the
visible and near-IR (Ibata \etal \cite{ibata99}, Egami \etal
\cite{egami00}, or in a section of an Einstein ring (Lewis \etal \cite{lewis02a}).
We estimate an overall
apparent size of the CO emission region from $u,v$-plane fits to the
CO(9--8) data.  A one-component elliptical Gaussian fit yields an
equivalent single-component CO size of $0^{\prime\prime}\!.6\pm 0^{\prime\prime}\!.2$ 
(Fig.\,\ref{visi-1mmlsb}).

\begin{table*}
\caption{Line Measurements in APM 08279+5255.}
\begin{tabular}{l c c c c c c c c}
\hline
Spectral line:
       &\multicolumn{2}{c}{--- CO(4--3) ---}    &CO(6--5)   
       &\multicolumn{2}{c}{--- CO(9--8) ---}    &CO(10--9)	  
       &CO(11--10)                              &HCN(5--4)\\      
Rest freq.(GHz)
       &\multicolumn{2}{c}{461.0408}            &691.4730
       &\multicolumn{2}{c}{1036.9120}           &1151.9850
       &1267.0145                               &443.1162\\
\hline
\multicolumn{5}{l}{ {\bf Observing Parameters:} }\\
Telescope &30\,m &PdBI  &30\,m   &30\,m   &PdBI    &30\,m   &30\,m  &PdBI \\
Observing freq.\ (GHz)
        &93.879 &93.867 &140.801 &211.141 &211.088 &234.573 &257.995 &90.229 \\
Beam FWHP (arcsec)
	&25 &$1.7\times 1.6$ &18 &11
	&$0.85\times 0.71$ &10.5 &9.5 &$7.2\times 5.4$\\
$S/T_a^*$ (Jy/K)
        &6.1    &---    &6.7     &7.9     &---     &8.7	    &9.5 &---\\
$T_{\rm mb}/S$ (K/Jy)
        &0.22   &47.8   &0.19    &0.23    &38.1    &0.20    &0.20 &2.90\\
Channel width (MHz)
        &1      &2.5    &4       &4       &2.5     &4      &4  &15 \\
Observing time (hr)
        &7.9    &26    &3.0     &5.2     &22.8    &4.8     &16.7 &27.9\\
\multicolumn{5}{l}{ {Rms noise in 50\,\kms\ :} }\\
1\,$\sigma$ in $T_a^*$ (mK)
        &0.3   &---    &0.5     & 0.6     &---     &0.5    &0.4 &--- \\
1\,$\sigma$ in $S$ (mJy)
        &1.9    &0.56   &3.1     &4.9     &1.3     &4.0     &4.1 &0.47 \\
\\
\hline
\multicolumn{5}{l}{ {\bf  Measured Line Parameters:} } \\
Line center (GHz)
	&93.869 &93.864    &140.794 & 211.137    &211.100 &234.616 &257.993 
	&90.210 \\
Redshift, $z$
	&3.9115 &3.9118    &3.9113  & 3.9111   &3.9119  &3.9110 &3.9110 
	&3.9121\\
Redshift error
        &$\pm 0.001$          &$\pm 0.0003$      &$\pm 0.0008$ & $\pm 0.0006$
	&$\pm 0.0003$          &$\pm 0.0015$      &$\pm 0.0015$ 
	&$\pm 0.0004$\
\\
Line peak ($T_a^*$, mK) 
        &1.2    &---    &2.1    &2.8      &---     &2.7     &2.2 &--- \\
Line peak  (mJy)         
        &7.3    &7.5    &14.3   &21.9     &24.2    &23.3    &20.5 &2.0 \\
Integ.flux (Jy\,\kms )
        &3.7    &3.8    &7.3    &11.1     &12.5    &11.9    &10.4 &0.85 \\ 
Line FWHM (\kms) &$490\pm60$ & $470\pm17$&$445\pm40$ &$480\pm40$ &$460\pm22$ &$480\pm90$ & $520\pm50$& $400\pm40$\\
\\
\hline
\multicolumn{8}{l}{ {\bf Apparent Line Luminosities (uncorrected for lensing):} }
\\
$L'_{\rm line}$ ($10^{10}\,L_l$)
        &14.7   &15.1   &12.7   &8.7      &9.8     &7.5     &5.4 &3.6 \\
$L_{\rm line}$ ($10^9$\,\Lsol )
        &0.46    &0.47    &1.3    &3.1      &3.5     &3.7     &3.5 &0.10 \\
\hline
\multicolumn{8}{l}{Errors in flux and luminosity are 
$\pm 10$\% for CO(4--3) and (6--5), 
and $\pm 20$\% for the other lines.}
\\  
\multicolumn{8}{l}{Line luminosity unit $L_l$ = \Kkmspc .}
\\
\multicolumn{8}{l}{Luminosities are for $H_{\rm 0} = 71$
\kms\,Mpc$^{-1}$, $\Omega_\Lambda=0.73$ and $\Omega_m=0.27$ (Spergel
\etal \cite{spergel03}), so at $z=3.91$, }  
\\
\multicolumn{8}{l}{angular diameter distance $D_A$ = 1.474\,Gpc, 
or 7147\,pc/arcsec, and luminosity distance $D_L$ = 35.556\,Gpc.}
\\
\label{linepara}
\end{tabular}
\end{table*}

\begin{table*}
\caption{Positions, Sizes, and Continuum flux densities.\label{tablepos}}
\begin{tabular}{lll ccccc}
\hline
 &R.A. 	  &Dec.	 &Major  &Minor	&P.A.  &Continuum	&Refs.\\
 &08$^{\rm h}$31$^{\rm m}$
	  &52$^\circ 45'$  &axis &axis &  &flux density
\\ 
{\bf Data} &(J2000)  &(J2000) &(arcsec) &(arcsec) &(deg.) &(mJy) \\	     
\hline
\multicolumn{5}{l}{ {\bf 1.4\,mm measurements:} }\\
1.4\,mm usb (dust)  
        &41$^{\rm s}.701$   
	&$17^{\prime\prime}\!.51$ 
	&$0^{\prime\prime}\!.57\pm 0^{\prime\prime}\!.05$
	&$0^{\prime\prime}\!.48\pm 0^{\prime\prime}\!.06$
	&$54^\circ\pm 25^\circ$
	&$16.9\pm 2.5$
	&1\\
1.4\,mm lsb (CO98+dust)	
        &41$^{\rm s}.702$   
	&$17^{\prime\prime}\!.49$ 
	&$0^{\prime\prime}\!.61 \pm 0^{\prime\prime}\!.03$
	&$0^{\prime\prime}\!.55\pm 0^{\prime\prime}\!.03$
	&$-14^\circ\pm 19^\circ$
        &---
	&1\\
CO(9--8) alone	
        &41$^{\rm s}.700$   
	&$17^{\prime\prime}\!.49$
	&$0^{\prime\prime}\!.72\pm 0^{\prime\prime}\!.08$
	&$0^{\prime\prime}\!.60\pm 0^{\prime\prime}\!.08$
	&$-25^\circ\pm 28^\circ$
        &---
	&1\\
\multicolumn{5}{l}{ {\bf 3.2\,mm measurements:} }\\
CO(4--3)
	&41$^{\rm s}.693$  
	&$17^{\prime\prime}\!.63$
	&$0^{\prime\prime}\!.89\pm 0^{\prime\prime}\!.1$
	&$0^{\prime\prime}\!.43\pm 0^{\prime\prime}\!.1$
	&$11^\circ\pm 10^\circ$
        &$1.3\pm 0.2$
	&1\\
HCN(5--4)
	&41$^{\rm s}.680$  
	&$17^{\prime\prime}\!.75$  &---    &---   &---  &$1.3\pm 0.2$   &2 \\
\multicolumn{5}{l}{ {\bf Optical and cm-wave measurements:} }\\
Optical astrometry:
        &41$^{\rm s}.68$
	&$17^{\prime\prime}\!.1$   &---   &---   &---  &---  &3 \\
HST NICMOS A image:  
        &41$^{\rm s}.64$  
	&$17^{\prime\prime}\!.5$   &---   &---   &---  &---  &4 \\
VLA--90\,cm & & & & &           &$5.1\pm 0.8$ &5  \\
VLA--20\,cm
        &41$^{\rm s}.708$  
	&$17^{\prime\prime}\!.44$   &---   &---   &--- &$3.05\pm 0.07$  &5 \\
VLA--3.6\,cm 
        &41$^{\rm s}.70$  
	&$17^{\prime\prime}\!.5$   &---   &---   &---  &$0.45\pm 0.03$  &4 \\
VLA--1.3\,cm  
        &41$^{\rm s}.70$  
	&$17^{\prime\prime}\!.5$   &---   &---   &---  &$0.41\pm 0.09$  &6 \\
\hline
 \multicolumn{8}{l}{{\it References:} (1) This paper; (2) Wagg \etal 2005 
and this paper; (3) Irwin et al.\ 1998 and private comm.; 
(4) Ibata et al. 1999;}
\\ 
\multicolumn{8}{l}{\ \ \ \ \ (5) Ivison 2006; (6) Lewis et al. 2002.}
\\  
\multicolumn{8}{l}{Estimated astrometric errors in the 1.4 and 3.2\,mm  
positions are $\pm 0.01^{\rm s}$ in R.A.\ and $\pm 0^{\prime\prime}\!.1$ in Dec.}
\\  
\end{tabular}
\end{table*}

\begin{figure}[hb] 
\centering
\includegraphics[angle=0,width=8.2cm]{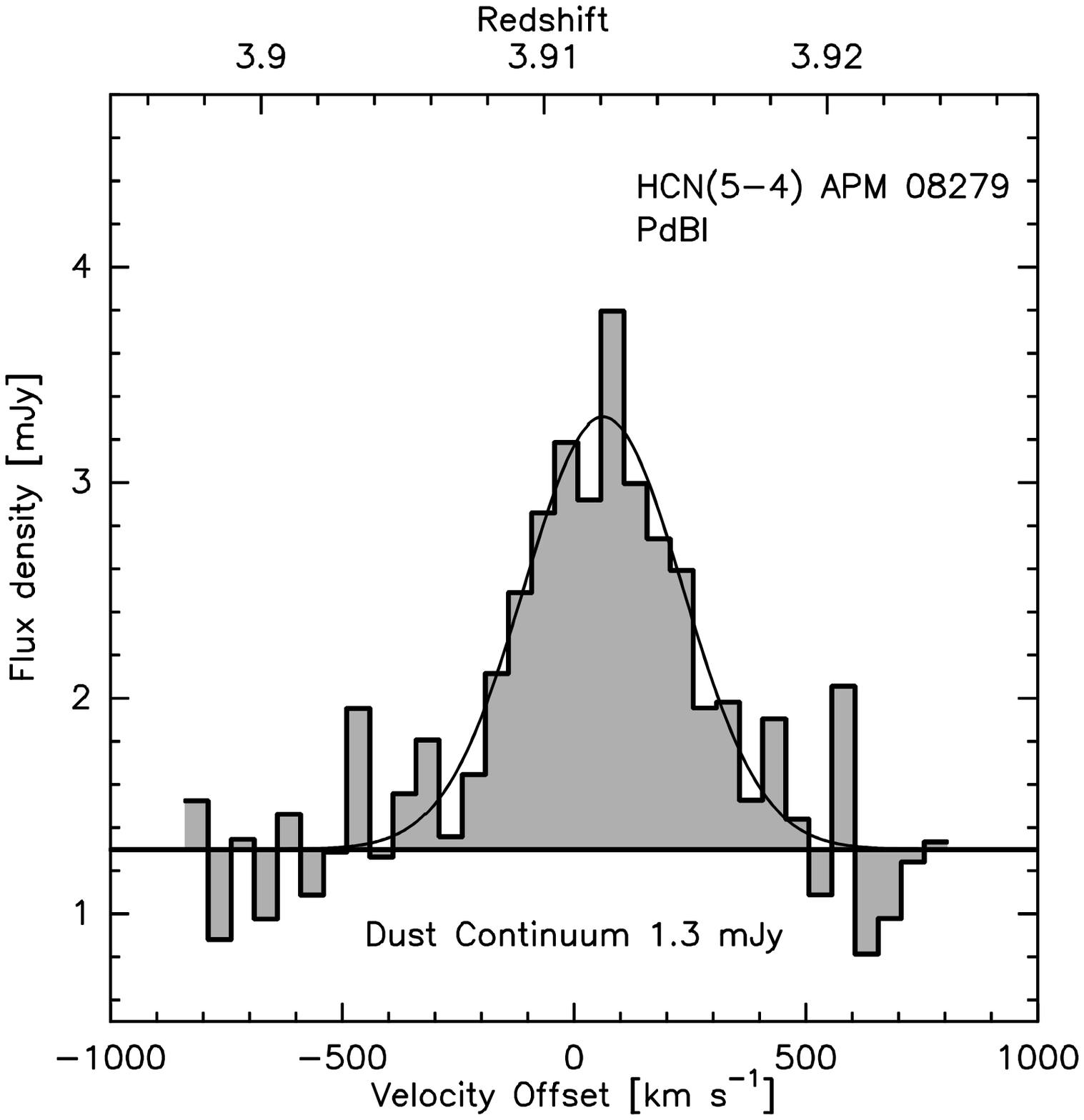}
\caption[New HCN54 Spectrum]
{
HCN(5--4) spectrum from APM\,08279+5255 obtained with the IRAM
Interferometer.  The line profile appears above the dust continuum
of 1.3\,mJy.  The velocity scale is relative to a redshift of
$z=3.911$, the beam is $7^{\prime\prime}\!.2 \times 5^{\prime\prime}\!.4$ at PA 81$^\circ$, 
and the channel width is 15\,MHz (49.84\,\kms ).  The rms noise
in the spectrum is 0.47\,mJy.  
}
\label{hcnline} 
\end{figure}

\subsubsection{HCN observations}
The first detection of the HCN(5--4) transition in APM\,08279 was recently 
reported by Wagg \etal (\cite{wagg05}).  
To better constrain the line parameters, in particular the line shape,  
we re-observed the HCN(5--4) line with the IRAM Interferometer's $D$
configuration in summer and autumn 2005, and also detected the HCN line in this new,
independent data set, which more than doubles the previous
on-source integration time. Our final spectrum (Fig.\ref{hcnline}) 
shows the combination of our new observations with the data published
by Wagg \etal (\cite{wagg05}), with the signal-to-noise ratio
improved by a factor of 1.4.  The Gaussian fit shown in
Fig.\,\ref{hcnline} yields a HCN(5--4) peak flux density of $2.0\pm
0.2$\,mJy, a velocity centroid of $+62\pm 15$\,\kms\ relative to our
tuning redshift of $z=3.911$ (90.229\,GHz), and an integrated HCN flux
density of $0.85\pm 0.10$\,Jy\,\kms, all in good agreement with the
values found by Wagg \etal.  
The dust continuum flux at this frequency is $1.3\pm
0.13$\,mJy. The measured HCN(5--4) linewidth of $402\pm
40$\,\kms\ is 15\% more narrow than the CO linewidth. This may indicate
that the HCN(5--4) line samples a slightly different region than the
CO lines.The difference of the linewidth, however, has only a 
low significance with $\Delta V_{\rm CO}-\Delta V_{\rm HCN}=75\pm45\kms$.

\section{Analysis}

\subsection{The dust continuum \label{dustcontinuum}}
\begin{figure*}
\centering
\includegraphics[width=16.0cm]{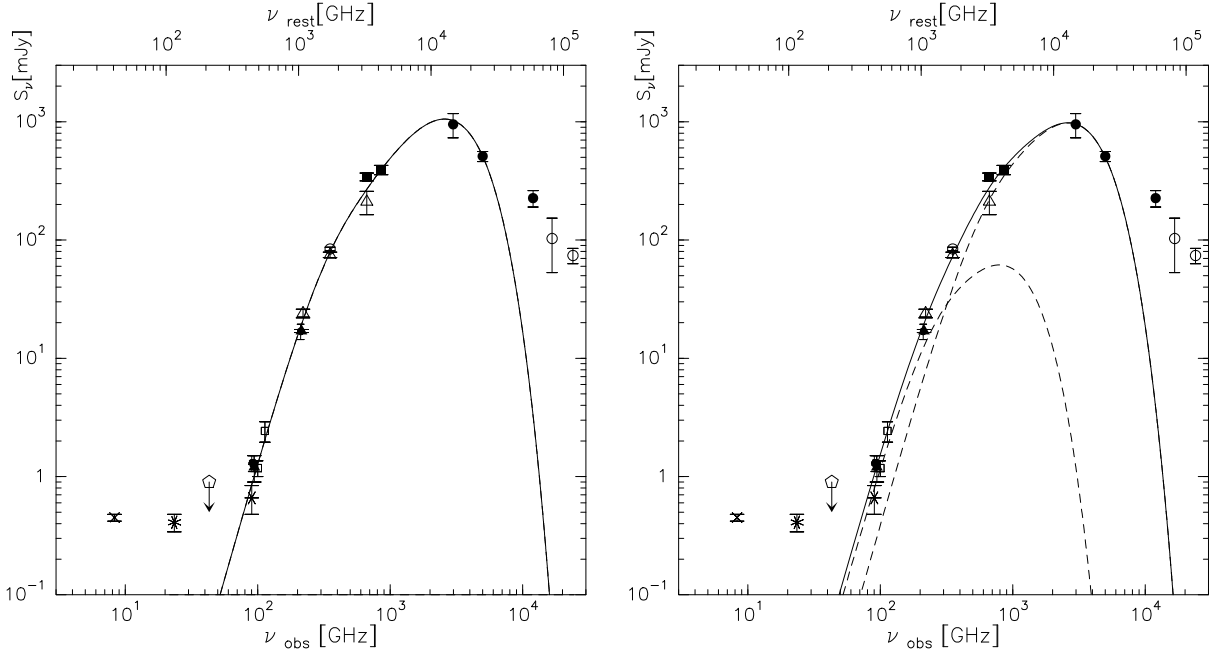}
\caption[Dust SED fits]
{Single component ({\it left}) and two component ({\it right}) dust models for APM 08279+5255.  
The continuum fluxes are from Irwin \etal
(\cite{irwin98}), Lewis \etal (\cite{lewis98}, \cite{lewis02a}), 
Downes \etal (\cite{downes99}), Egami \etal (\cite{egami00}), Papadopoulos \etal (\cite{papado01}),
Barvainis \& Ivison (\cite{barvainis02}), Wagg \etal (\cite{wagg05}), Beelen \etal
(\cite{beelen06}) and this work.
}
\label{dustseds} 
\end{figure*}

We use our new continuum measurements together with data between
3\,mm and 60\microns\ (restframe 700--14\microns) from the literature
(for references see caption of Fig.\,\ref{dustseds}) to analyze the dust
temperature and the size of the dust emission region.  Because the dust
emission in ULIRGs becomes optically thick at a rest wavelength of  
$\sim 100$\microns\ (Downes \etal \cite{downes93}) we did not use the
optically thin approximation, but
\begin{equation}
S_{\nu}=\bigg( B_{\nu}(T_{\rm dust})-B_{\nu}(T_{\rm
BG})\bigg){{(1-e^{-\tau_{\nu}}) }\over {(z+1)^3} } \Omega_{\rm app} \
\ , 
\end{equation}
where $B_{\nu}$ is the Planck function.
Note that since $S_{\nu}$ is the {\it observed} (amplified) flux density, 
the lens magnification, $m$, is hidden in the apparent solid angle, $\Omega_{\rm app}$.  
The dust optical depth, however, is a property of the source itself,
and is independent of the gravitational lensing:
\begin{equation}
\tau_{\nu}= \kappa_{\rm d}(\nu_{\rm r})\,M_{\rm
dust,\,app}/( D_{A}^2\,\Omega_{\rm app})  \ \ \ ,
\end{equation}
where $\kappa_{\rm d}$ the dust absorption coefficient,
$M_{\rm dust, app}$ the apparent dust mass, and $D_{A}$ is the angular diameter 
distance to the source (here the magnification cancels out via $M_{\rm
dust,\,app}/\Omega_{\rm app}$). For the frequency dependence of the dust absorption
coefficient we adopt
\begin{equation}
 \label{dustfreqdependence}
\kappa_{\rm d}(\nu_{\rm r})\ 
=\ 0.4\,(\nu_{\rm r}/250\,{\rm GHz})^\beta\,
\end{equation}
with units of cm$^{2}$ per gram of dust
(Kr\"ugel \& Siebenmorgen \cite{kruegel94}), and with $\beta= 2$ (Priddey
\& McMahon \cite{priddey01}).  Following our CO analysis, we
substitute $\Omega_{\rm app}$ by the apparent equivalent radius
$r_{0}$ so that the dust spectrum can be described by $T_{\rm dust}$,
$M_{\rm dust, app}$, and $r_{0}$.

\noindent
The continuum data is equally well fit with a single or 2-component
dust model. The single component fit is shown in
Fig.\,\ref{dustseds} (left). From this fit we find $T_{\rm dust} \approx
215\pm 10$\,K, $M_{\rm dust}\approx$ $7.5\cdot
10^8\,m^{-1}$\,\msol\,($\pm10$\%) and  $r_0 \approx 680\pm 35$\,pc.
The 2-component fit is shown in Fig.\,\ref{dustseds} (right). For convenience 
we call these the ``cold'' and ``warm''
components.  They correspond to the ``starburst'' and ``AGN'' components
respectively, in similar fits to the dust 
spectrum by Rowan-Robinson (\cite{rowan00}) and Beelen \etal (\cite{beelen06}; see also
the multi-component fits to APM\,08279 by Blain \etal \cite{blain03}).
To reduce the number of free parameters, we assume that the overall size
of the dust emission region is similar to that of the CO lines, and
fix the equivalent (magnified) radius $r_0$ to 1150\,pc (from the 2-component CO
model).  We then determine dust temperatures and
masses (times magnification $m$) of both components, and the relative
area filling factor of the warm and cold components.  We fit these five
free parameters to the data points above 43\,GHz (upper limit only)
and below the IRAS 25\microns\ point. 
%
%
\noindent For the ``cold'' dust component, 
we find $T_{\rm cold} \approx 65\pm 18$\,K and
$M$(cold dust) $\approx$ $2.6\cdot 10^9\,m^{-1}$\,\msol\,($\pm25$\%).
The ``warm'' component has $T_{\rm warm}$ $\approx$  220$\pm$30\,K, 
$M$(warm dust) $\approx$ $2.1\cdot 10^8\,m^{-1}$\,\msol\,($\pm$65\%). 
The area filling factors we derive for 
the ``cold'' and ``warm'' dust components are
70$\pm$10\% and 30$\pm$10\% respectively. 
For the ``cold'' component, the dust continuum becomes opaque for 
rest wavelengths shorter than 300\,\microns\ ($\lambda_{\rm
obs} \approx 1.3$\,mm). 
The implied FIR luminosity is $L_{\rm FIR}$ $\approx$ 
$2.0\cdot 10^{14}\,m^{-1}\,\lsol$
(with $L_{\rm FIR}$ integrated from 40--120 \microns\ restwavelength,
Helou \etal \cite{helou85}).
The contributions from the cold and warm components are
$2\cdot 10^{13}\,m^{-1}\,\lsol$ and
$1.7\cdot 10^{14}\,m^{-1}\,\lsol$ respectively.

\subsection{The CO line SED}

The peak and velocity-integrated fluxes of the CO lines (Table
\ref{linepara}) increase with rotational quantum number up to the 
9--8 line. Beyond this transition the CO line SED flattens or even
starts to decrease as for the 11-10 line. The peak of the CO line
SED occurs at $10-9$ line.   
The CO line luminosity, $L^\prime_{CO}$, in \Kkmspc , is defined by 
\begin{equation}
  L^\prime_{\rm CO} = 3.25\cdot 10^7\ S_{\rm CO}\,\Delta V \nu^{-2}_{\rm obs}\,
D^2_L\,(1+z)^{-3}\ \ \ ,
\end{equation}
where the integrated CO flux $S_{\rm CO}\,\Delta V$ is in Jy\,\kms , 
the observed frequency $\nu_{\rm obs}$ is in GHz, and the luminosity
distance $D_L$ is in Mpc (e.g., Solomon \etal \cite{solomon97}).
For transitions with $J>4$ the CO line luminosities in APM\,08279 
are monotonically decreasing which shows that theses lines are
subthermally excited. Interestingly, the
line luminosities of the CO(1--0) and CO(2--1) transitions are also
lower than the line luminosity of the CO(4--3) line (Papadopoulos
\etal \cite{papado01}; Riechers \etal
\cite{riechers06}), which indicates that the low-$J$ transitions of 
CO have only moderate opacities.

\subsubsection{CO LVG Modeling}
To investigate the CO excitation in more detail, we first apply a
spherical, single-component, large velocity gradient (LVG) model. We
use the collision rates from Flower (\cite{flower01}) with an \hh\
ortho-to-para ratio of 3.  To compare the observations with the
LVG-predicted Rayleigh-Jeans brightness temperatures, $T_{b}$, 
we convert the latter to flux densities via
\begin{equation}
\label{lvgladder}
S_{\rm CO} = \Omega_{\rm app}\,T_{b}/(1+z)\,2k\nu_{\rm obs}^2/c^2 \ ,
\end{equation}
where $\Omega_{\rm app}$ is the apparent source solid angle (i.e.,
true source solid angle times gravitational magnification). Our
analysis thus yields not only the gas density and temperature, but
also an estimate of the magnified CO size that can be compared with
the interferometer measurements. Note that $\Omega_{\rm app}$ is not
a function of the rotational quantum number $J$ for a given 
LVG solution. We define the equivalent source radius as
\begin{equation}
r_{0}= D_{A}\,\sqrt{\Omega_{\rm app}/\pi} \ ,
\end{equation}
where
$D_{A}$ is the angular size distance to the source, and $\Omega_{\rm
app}$ is the magnified solid angle that we derive from the LVG-model
brightness temperatures.  This equivalent radius $r_{0}$ would be
equal to the true source radius if the CO were in an unlensed, 
face-on, filled circular disk.

\noindent The total \hh\ mass from the LVG models can be obtained using the CO
emitting area (expressed as $r_0$) and the \hh\ column density calculated via
\begin{equation}
N_{\hh}= 3.086\cdot 10^{18}\,n(\hh)\,{{\Delta\,V_{\rm turb}} \over  {dv/dr} }
\end{equation}
to estimate the number of \hh\ molecules and therefore the gas mass.
In this equation $\Delta\,V_{\rm turb}$ is the turbulence linewidth and $dv/dr$ the velocity gradient
of the LVG model. Thus $\Delta\,V_{\rm turb}/\gradv$\ is the equivalent path length
through the molecular disk. The LVG gas mass is given by 
\begin{equation}
\label{lvgmasseq}
M_{\rm LVG}= 0.21\,r_0^2\,n(\hh)\,{{\Delta\,V_{\rm turb}} \over  {dv/dr} }\ [\msol].
\end{equation}
This LVG mass includes a correction of 1.36 to account for the mass contribution of Helium.
Note that $\Delta\,V_{\rm turb}$ in APM\,08279 
is not equal to the observed linewidth as the later is dominated by the
rotation of the circumnuclear disk and not by its radiatively
effective random motions of the gas along a line of sight.

\subsubsection{Single-component LVG models for CO}

We first use a CO abundance per velocity
gradient of [CO]/$\gradv = 8\cdot 10^{-5}\,{\rm pc}\,(\kms)^{-1}$.
For this value, models with enough excitation to match the observed 
high-$J$ fluxes, however, predict high CO opacities, even for
CO(1--0),  so the line luminosities are predicted to stay constant from 
CO(1--0) to (4--3), while the observations show a deficit in the lowest
lines (Papadopoulos \etal \cite{papado01}). For lower CO abundance per velocity gradient, the CO opacities 
are lower, because the implied CO column densities per velocity interval are
reduced. For [CO]/$\gradv = 1\cdot 10^{-5}\,{\rm pc}\,(\kms)^{-1}$,
most of the LVG models that fit the high-$J$ lines also have low enough
opacities to reproduce the observed fluxes of 
the low-$J$ CO lines. Thus all the lines from CO(1--0) to CO(11--10) could be fit with a
{\it single-component} LVG model. Similar low values for [CO]/$\gradv$ (i.e, a
large velocity spread) in nearby starburst galaxies likely also
explain the low $^{13}$CO/$^{12}$CO ratios observed in their nuclei
 (e.g. Aalto \etal \cite{aalto91}, Paglione \etal \cite{paglione01}). 
In the following, we therefore fix [CO]/$\gradv$ to $1\cdot 10^{-5}\,{\rm
pc}\,(\kms)^{-1}$.

\begin{figure}[h] 
\centering
\includegraphics[width=8.5cm]{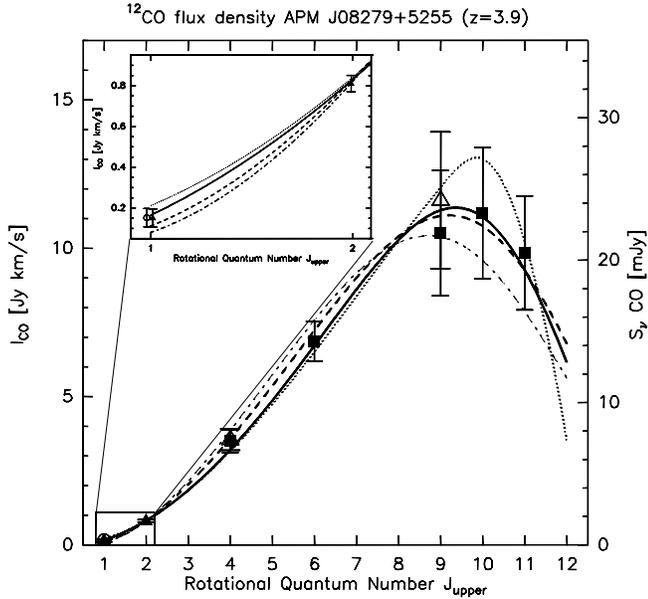}
\caption[CO SED]
{Observed CO fluxes vs.\ rotational quantum number (CO line SED) for
APM\,08279+5255, obtained with the IRAM 30\,m telescope (filled
squares) and the IRAM Interferometer (open triangles) (D99; this
paper). Errors include the calibration uncertainties. The fluxes for the 1--0 and 2--1 lines are from Papadopoulos
\etal (\cite{papado01}, filled triangles), and Riechers \etal
(\cite{riechers06}, circle at $J$=1). 
The single-component LVG model fluxes are shown for ($n(\hh)$, $\tkin$) 
combinations as follows: 
solid line: ($10^{4.4}$ cm$^{-3}$, 125\,K);
dashed line: ($10^{4.2}$ cm$^{-3}$, 220\,K);
dotted line: ($10^{5.4}$ cm$^{-3}$, 40\,K);
dashed-dotted line: ($10^{4.0}$ cm$^{-3}$, 350\,K).
The inset shows a zoom, for the observed and model-predicted fluxes of 
the 1--0 and 2--1 lines. The two CO(1--0) data points are shown with
a small offset in the  rotational quantum number for legibility reasons.
}
\label{cosed} 
\end{figure}

\begin{figure} \centering
\includegraphics[width=8.0cm]{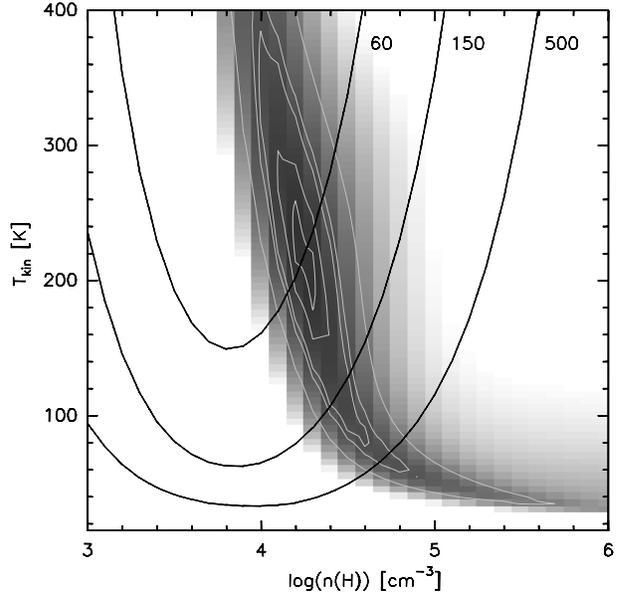}
\caption[$\chi^2$ distribution]
{$\chi^2$ distribution for a single component LVG model fit to the
observed line luminosity ratios (greyscale and grey contours,
contours: $\chi^2=2$, 4, 8, 10 and 20).
The CO abundance per velocity gradient for the LVG models is [CO]/$\gradv =1\cdot 10^{-5}\,{\rm
pc}\,(\kms)^{-1}$. Black lines show gas to dust mass
ratios of 60, 150 and 500 calculated from the LVG \hh\ mass for a dust mass of $M_{\rm dust}=7.5\cdot
10^8\,m^{-1}$\,\msol (single component dust fit, see
Sec.\,\ref{dustcontinuum}). For the calculation of the LVG \hh\ mass
we have used a turbulence line width of $dv_{\rm turb}=100\,\kms$ and
a velocity gradient of $\gradv=5\,\kms\,{\rm pc}^{-1}$ (see also Sec.\,\ref{gasmass})}.
\label{chisq-1comp} 
\end{figure}

\noindent Fig.\,\ref{cosed} shows the results of fitting the observed
CO fluxes with selected single-component LVG models.  A good fit to
the observations is provided by an H$_2$ density of 10$^{4.4}$
cm$^{-3}$, a gas temperature of $\tkin=125$\,K and an equivalent
(magnified) radius of $r_0=910$\,pc. For these parameters, the
CO(1--0) optical depth is only 1.3.  The CO(2--1)/CO(1--0) line
luminosity ratio is then 1.25, in agreement with the observations
(Papadopoulos \etal \cite{papado01}). The best fitting solution is
provided by $n$(\hh) = 10$^{4.2}$ cm$^{-3}$, $\tkin=220$\,K and
$r_0=790$\,pc.  Other temperature--density combinations with similar
\hh\ pressure also match the data (Fig.\,\ref{cosed}, Fig.\,\ref{chisq-1comp}).
 For \hh\ densities below 10$^{4.0}$ cm$^{-3}$,
however, the model predictions fail
to reproduce the observed fluxes in the high-$J$ lines. 
The kinetic temperature is poorly constrained by our
models. Lower-temperature 
models with H$_2$ densities greater than 10$^{5}$ cm$^{-3}$
and $\tkin \sim 50$\,K will also make the CO line SED turn over at
the 10--9 line. For these cold, dense solutions, the turnover
of the CO SED is much steeper than for the warm solutions, so
in case better data were to become available, this predicted steep 
slope might allow us to distinguish between cold and warm gas. 
For the densest solutions, however, the models 
overestimate the 1--0 flux because the low-$J$ opacities 
become too high. In Fig.\,\ref{chisq-1comp} we show the
results of a $\chi^2$ test for the single component LVG model. In this
plot we also show the resulting gas to dust mass ratio. The plot
demonstrates that solutions with an \hh\ density in excess of $10^5\,{\rm
cm}^{-3}$ lead to very high LVG gas masses which imply
unrealistically high gas to dust mass ratios ($>1000$). The warmest and coolest possible LVG solutions
limit the equivalent (magnified) size of the emission region to the range 
$r_0\approx750-1850$\,pc. Using a gas to dust mass ratio of 500 as a plausible upper limit
restricts the emission region further to $r_0\approx750-1350$\,pc.

\subsubsection{Two component LVG model for CO \label{cotwocomp}} 
\begin{figure}[h] 
\centering
\includegraphics[width=8.5cm]{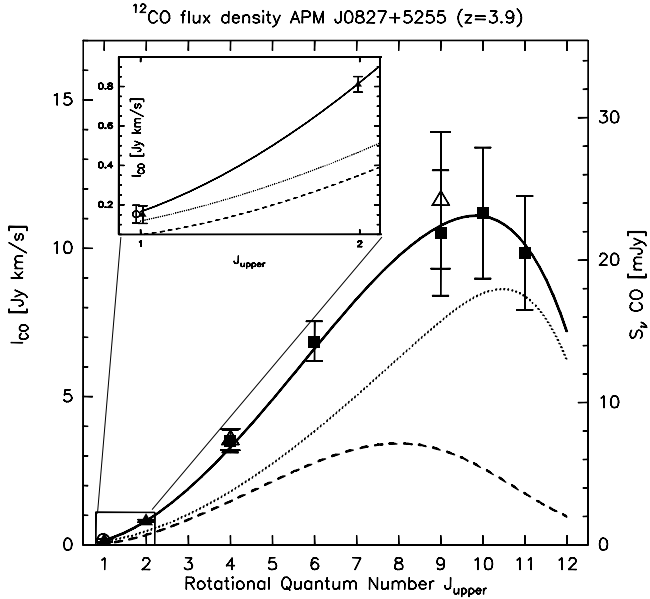}
\caption[2-component model]
{2-component model for the CO lines.  
The dotted line represents the ``cold'', dense gas, the dashed line
the ``warm'' gas and the solid line the sum of both components.  Model parameters
are listed in Table~3.  
}
\label{co2comp} 
\end{figure}

Similar to the dust SED we can also split the CO line SED into two components. 
Such a 2-component model matches the widespread 
view that the ISM contains dense, star-forming molecular cloud cores 
(traced by HCN) and more diffuse gas in the cloud envelopes
(e.g. Solomon \etal \cite{solomon92}). In this picture, 
the dense HCN-emitting cores also emit CO lines, with a CO-SED
extending to higher $J$ than the CO-SED of the lower-density clouds. We therefore fit 
the CO-SED with a 2-component model, so that the sum of {\it both} CO 
components agrees with the observed CO SED.
As the kinetic temperature for both components 
is expected to be similar to, or higher than the dust temperature we have used the dust temperatures to 
break the degeneracy between the kinetic temperature and the \hh\ density. A good solution for all the observed 
CO lines with kinetic temperatures similar to the dust temperatures is shown in
Fig.\,\ref{co2comp}. This model has a high-density component
with $n$(\hh)$\approx 10^{5.0}$\,cm$^{-3}$, $\tkin \approx 65$\,K, and a
source equivalent (magnified) radius of $r_0\approx995$\,pc. 
It also has a warmer, lower-density component 
with $n$(\hh)$\approx 10^{4.0}$\,cm$^{-3}$, $\tkin \approx 220$\,K, 
and $r_0\approx575$\,pc.

%

This 2-component model does not necessarily imply that the warmer gas is
closer to the AGN.  The two components could just as well be randomly
mixed together in a single circumnuclear disk with an equivalent
(magnified) radius of $r_0\approx1150$\,pc, with relative area filling
factors of 75\% and 25\% for the cooler and warmer gas
respectively - similar to the relative filling factors derived from the dust continuum.
In contrast to the single component model the dense gas phase in this 2-component model
only results in a gas to dust mass ratio of $\sim 150$ as the dust mass in the
cold dense component is much higher than the dust mass derived from a single component
fit. Interestingly, the contribution of the cold, dense gas to the CO(1--0) line luminosity
exceeds that of the warm gas phase. The (gravitationally-amplified) CO(1--0) line luminosities 
associated with both components are $L^\prime_{\rm CO}$(cold) = $7.4\cdot 10^{10}$\,\Kkmspc\ 
and $L^\prime_{\rm CO}$(warm) = $3.0\cdot 10^{10}$\,\Kkmspc.

\subsection{Relation to HCN}
\subsubsection{Collisional Excitation}
Our finding that the CO emission arises from high-density gas
($n(\hh)>10^{4}$ cm$^{-3}$) is supported
by the detection of the HCN(5--4) line in this source (Wagg \etal
\cite{wagg05}). In the following we use our CO single component LVG
solutions to check whether the HCN(5--4) emission emerges from the
same CO-emitting volume, or if we need additional gas at even higher
density, {\bf as in the 2-component CO model}, to explain the observations.
For the LVG models of HCN, we use collision rates from
Schoier \etal (\cite{schoier05}) and assume a relative abundance
ratio of [HCN/CO] = $10^{-4}$ (Helfer \& Blitz \cite{helfer97}; Wang
\etal \cite{wang04}). This gives [HCN]/$\gradv=1\cdot 10^{-9}\,{\rm pc}\,(\kms)^{-1}$.  
The net result is that {for the single component model} only the coldest, densest models 
($n$(\hh)$\approx 10^{5.4}$ cm$^{-3}$, $\tkin \approx 40$\,K,
$r_0\approx 1800$\,pc) can excite HCN to
produce the observed HCN(5--4) flux density.  As noted above, these
models fail to reproduce a CO(2--1)/CO(1--0) ratio larger than unity, lead to
unrealistically high gas to dust mass ratios and are also inconsistent with the high dust temperatures derived for
APM\,08279.  Conversely, warmer CO models with lower \hh\
density provide less than 15\% of the observed HCN(5--4) luminosity.
Hence, a single-component model cannot simultaneously
explain the CO and HCN line luminosities.  

\begin{figure*}[ht] 
\centering
\includegraphics[width=18.0 cm]{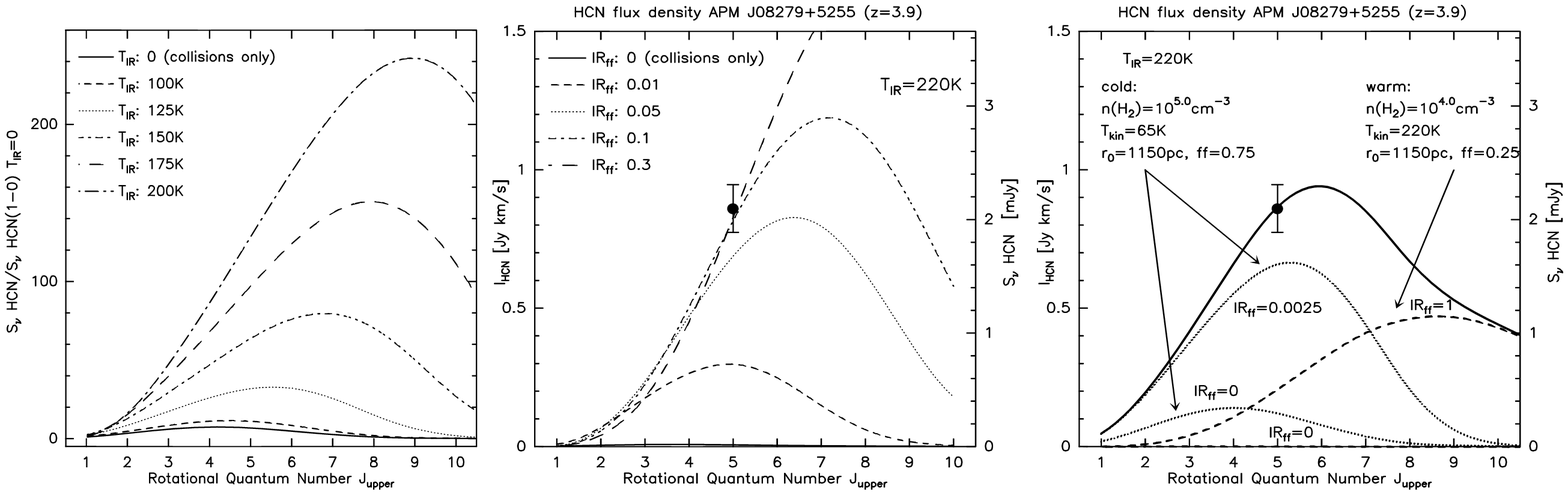}
\caption[IR pumping for HCN]
{{\it Left:} HCN line SED as a function of the IR radiation temperature for a \hh\ density 
of $10^{5}$ cm$^{-3}$, a kinetic temperature of  $65$\,K and an delution factor for the IR
radiation field of 10\% for each model. All SEDs are normalized to the HCN(1--0) flux density
for the pure collisional excitation model ($T_{\rm IR}=0$\,K). The plot exemplifies the large
effect of IR pumping on the HCN excitation for dust temperatures above $\sim100$\,K. {\it Middle:}
HCN excitation in APM\,08279+5255 for gas parameters as derived from the single component dust/CO model 
( $n$(\hh) = 10$^{4.2}$ cm$^{-3}$, $T_{\rm IR}=\tkin=220$\,K and $r_0=790$\,pc) as a function of
the IR filling factor. For IR$_{\rm ff}=10$\% or above the IR pumping boosts the HCN excitation 
such that the HCN(5--4) flux density is in agreement with the observations. {\it Right:}
HCN excitation model for the two--component dust/CO model for APM 08279+5255.  
The dashed line represents the ``warm'' gas with and IR illumination factor of unity, the dotted lines 
the ``cold'', dense gas with pure collisional excitation (IR$_{\rm ff}=0$) and with an IR illumination of 0.25\%.
The solid line the sum of both IR illuminated components. The IR radiation temperature for both components is 220\,K.
}
\label{hcn-ir-pumping} 
\end{figure*}

\noindent Wagg \etal (\cite{wagg05}) have argued that an increased HCN abundance
may be responsible for the bright HCN emission. Their calculations suggest that 
a relative HCN abundance of [HCN/CO]\,$\approx 10^{-2}$, thus 2 orders of magnitude
above the standard abundance ratio (Wang \etal \cite{wang04}), is
required in order explain the observed HCN and CO luminosities from a
single gas component. Enhanced HCN abundances can be caused by an
increased ionization rate in the vicinity of an AGN (see Wagg \etal
\cite{wagg05} and reference therein). Models however suggest that this
increase is only modest ([HCN/CO]\,=\,$5\cdot10^{-4}$, Lepp \&
Dalgarno \cite{lepp96}). For NGC\,1068 Usero \etal (\cite{usero04})
suggested that X-rays drive the abundances ratio to
[HCN/CO]\,$\approx10^{-3}$. But even if we use this increased HCN
abundance, single component models with kinetic temperatures similar to
the dust temperature fail to reproduce the observed HCN(5--4) flux
density by a factor of 2 or more.

\noindent
In the context of the 2-component model the predicted HCN(5--4) luminosity
increases because the density of the cold dense gas is much higher than that derived
from the one--component model. But even for this gas phase,
which has a density comparable to that derived for the HCN emitting
volume in local IR luminous sources (Greve \etal \cite{greve06}), the predicted 
HCN(5--4) luminosity is much smaller than the observed value. 
We can, however, increase the \hh\ density of the cold gas component if kinetic
temperatures below the dust temperature are considered. An equally good fit to 
all the observed CO lines that also matches the HCN(5--4) luminosity can 
be obtained using a high-density component
with $n$(\hh)$\approx 10^{5.7}$\,cm$^{-3}$, $\tkin \approx 45$\,K, and a
source equivalent (magnified) radius of $r_0\approx1125$\,pc as well
as a warmer, lower-density component  with $n$(\hh)$\approx 
10^{4.0}$\,cm$^{-3}$, $\tkin \approx 200$\,K, and $r_0\approx650$\,pc.
The high density for the cold gas component, however, leads again to
a very high gas to dust mass ratio of $\sim 1000$. Given the high metalicity
of APM\,08279 (2 to 5 times solar as
suggested by X-ray observations of the iron K-shell absorption edge, 
Hasinger \etal \cite{hasinger02}), this makes it very unlikely 
that the HCN(5--4) luminosity can be explained within a pure collisional excitation scheme.

\subsubsection{IR-pumping of HCN \label{irpump}}

\noindent Another potential mechanism to increase the HCN luminosity is 
to boost the HCN by infrared excitation, notably through the 
stretching and bending modes at 3, 5, and 14\,$\mu$m. 
Various studies have addressed this issue for local IR bright galaxies
and concluded that IR pumping via the 14\,$\mu$m modes is not 
important compared to collisional excitation (e.g. Stutzki \etal \cite{stutzki88}, 
Paglione, Jackson \& Ishizuki \cite{paglione97}, Gao \& Solomon \cite{gao04}).
The dust SED of APM\,08279, however, has a much higher contribution of warm dust
compared to other sources studied in detail so far. Furthermore its dust temperature of $\sim 200$\,K is
well beyond the minimum dust temperature of $\sim\,110$\,K required for IR pumping via
the 14\,$\mu$m bending mode to become efficient (Carroll \& Goldsmith \cite{carroll81}).\\

\noindent To investigate the effect of the increased IR field in APM\,08279+5255
on the HCN excitation we include the first vibrational
bending mode ($\nu_2=1$, see e.g. Thorwirth \etal \cite{thorwirth03}) in our LVG code.
We neglected the vibrational $\nu_1$ mode as its excitation requires $\sim 4$ times higher
IR temperatures (Carroll \& Goldsmith \cite{carroll81}).
For the computation we take 20 rotational levels into account which
leads in total to 58 energy levels due to the $l-type\ doubling$ of the $\nu_2=1$ mode. 
The IR field is described by a greybody spectrum with a frequency dependence 
in analogy to Eq.\,\ref{dustfreqdependence}. 
The illumination of the dense gas by the IR field is parameterized by an
IR delution factor. This filling factor represents the solid angle fraction of the gas exposed
to the IR field.\\
Fig.\,\ref{hcn-ir-pumping} (left) exemplifies the variations of the HCN
line SED as a function of the IR radiation temperature. We find that for 
dust temperatures below 100\,K the excitation through vibrational pumping 
is small compared to the collisional excitation for \hh\ densities typical for
HCN emitting regions ($\sim 10^{4-5}$\,cm$^{-3}$). This is in line with the 
finding that IR pumping is not an important mechanism for local IR bright galaxies 
and most high-$z$ sources as their dust SEDs show a much smaller contribution
of dust with temperatures in excess of $\sim 50$\,K compared to APM\,08279  
(e.g. Lisenfeld \etal \cite{lisenfeld00}, Beelen \etal \cite{beelen06}).
For increasing dust temperatures, however, the IR pumping becomes very efficient.
At a dust temperature of about 200\,K the HCN excitation is completely
dominated by IR pumping as long as the IR delution factor is above a
few percent.

\noindent In Fig.\,\ref{hcn-ir-pumping} (middle) we show the effect of the IR dust field on the 
HCN excitation for the 1--component CO model ($T_{\rm IR}=210$\,K, $\nhh=10^{4.2}$
cm$^{-3}$, $\tkin=220$\,K, $r_0=790$\,pc).
For this model the collisional excitation of the HCN(5--4) line is
negligible. But already an IR delution factor of $\sim5\%$ boosts the
HCN luminosity by almost 2 orders of magnitude. For a filling
factor of 10\% or higher the IR pumping is strong enough to explain
the observed HCN(5--4) luminosity in APM\,08279. Thus the exceptional high
dust temperature in APM\,08279 provides a straightforward
explanation for the strong HCN luminosity in this source without
any ad hoc changes of the HCN abundance or extreme \hh\ densities. This most 
likely also holds for the HCO$^+$ emission recently detected in APM\,08279 
(see Garcia-Burillo \etal \cite{garcia06} and there discussion on the IR 
pumping of HCO$^+$).

\noindent Although the single component model with IR pumping yields a good fit 
to the CO and HCN line intensities as well as to the dust continuum it is 
presumably too simple as the HCN emission is known to arise from cloud cores 
with are typically an order of magnitude denser than those traced by the bulk of the
CO emission. We therefore also show the effect of IR pumping for the 2-component CO model 
(Fig.\,\ref{hcn-ir-pumping}, right). Since the IR filling factors for both components can
not be derived from the available data, we consider here only the extreme case where the
IR filling factor for the warm component, which is responsible for the dust emission at 220\,K, is unity. 
From the Figure it can be seen that the IR pumping of the warm component, even in this extreme case, 
is not sufficient to explain the observed HCN(5--4) luminosity. Its contribution to 
the HCN(5--4) flux is similar to that of the pure collisional excitation from the 
cold dense medium. But even the sum of both mechanisms still underpredicts the HCN(5--4) luminosity
so that also a small part of the cold dense gas needs to be exposed to the 220\,K dust field. 
The required IR filling factor for the cold dense gas, however, is only 0.25\%. 
The HCN(1--0) line luminosity from this model is dominated by the emission from the 
cold, dense gas and yields $L^\prime_{\rm HCN} \approx 5\cdot 10^{10}$\,\Kkmspc. The contribution
due to collisional excitation is $L^\prime_{\rm HCN}\approx 2\cdot 10^{10}$\,\Kkmspc. We note
that the IR filling factor for the cold gas remains low even if the IR filling for the warm gas
is smaller than unity. That is, in the absence of the warm gas component the IR filling for the cold
gas required to match the observed HCN(5--4) luminosity is still below 1\%.

\subsection{Magnification and size of the Molecular Gas distribution \label{magsection}}

In the following we use our effective (magnified) radius 
$r_0$ derived from the LVG and dust models in combination with the
radial dependence of the magnification from the lens model 
of Egami \etal (\cite{egami00} Fig.\,9)
to re-estimate the magnification and intrinsic size of the CO 
and dust emission regions.  By definition,    
the apparent CO luminosity, $L^\prime_{\rm CO}$, is related to 
the effective radius $r_0$ by 
\begin{equation} 
L^\prime_{\rm CO} = T_b\,\Delta V\,\pi\, r_0^2  \ \ \ , 
\end{equation}
where $T_b$ is the brightness temperature of the CO line and $\Delta
V$ the observed line width.
For a face--on, filled and isothermal disk we can express the radial dependence of the 
integrated CO luminosity in the case of radial differential
magnification by
\begin{equation} 
L^\prime_{\rm CO}(r) = T_b\,\Delta V\,2\pi \int\limits_{0}^{r} m(r')\,r'\,dr'   \ \ . 
\end{equation}
Substituting $T_b\,\Delta V$ by the 
observed CO luminosity and its equivalent radius and taking into account that the CO can not survive
closer to the central black hole than a minimum radius, $r_{\rm min}$, thus allows us to 
express $L^\prime_{\rm CO}$ as a function of the 
intrinsic radius via
\begin{equation} 
\label{lcoface}
L^\prime_{\rm CO}(r) = \frac{L^\prime_{\rm CO\,obs}}{\pi r_0^2}\,2\pi
\int\limits_{r_{\rm min}}^{r}m(r')\,r'\,dr'   \ \ \ . 
\end{equation}
In the following we assume $r_{\rm min}=1$\,pc which corresponds to the radius where the temperature 
for gas heated by the AGN with an intrinsic luminosity of $L_{\rm bol}=5\cdot10^{13}\,\lsol$ (Egami \etal \cite{egami00}) 
has dropped below the dust sublimation temperature of $\sim 1500$\,K.
We can now compare the radial dependence of the  $L^\prime_{\rm CO}$ to the observed CO 
luminosity to derive the true source radius $r_{\rm true}$.
The effective magnification can then be calculated via
\begin{equation} 
m_{\rm eff} = \frac{r_0^2}{r_{\rm true}^2-r_{\rm min}^2} \ \ .
\end{equation}

\begin{figure}[h] 
\centering
\includegraphics[width=8.5cm]{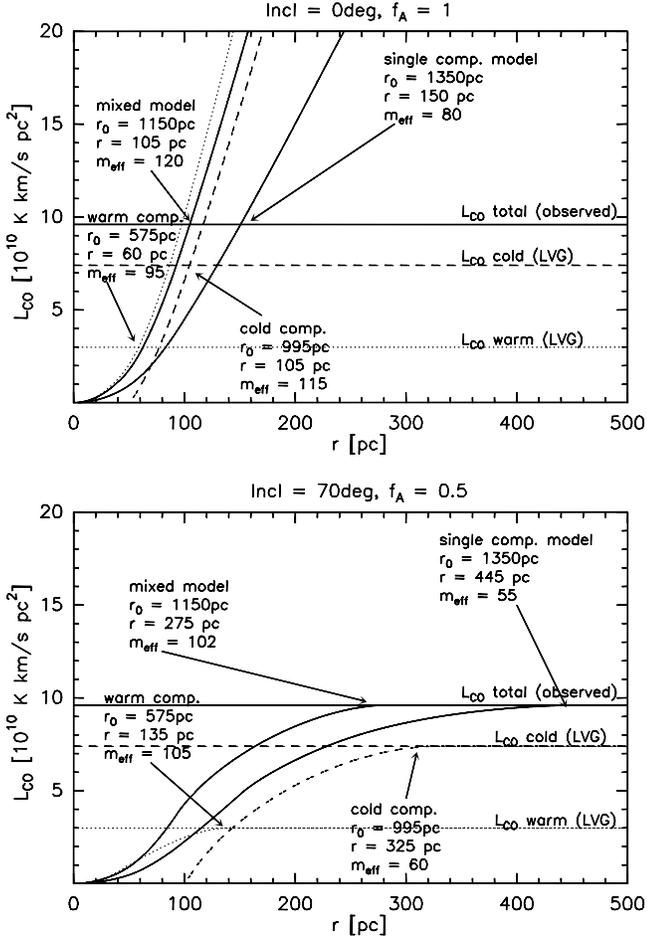}
\caption[LCO-radius Relation]
{Radial dependence of $L^\prime_{\rm CO}$ calculated using the 
differential triple-image lens model from Egami \etal
(\cite{egami00}, their Fig.\,9) for a filled, face--on disk (top) and a disk
seen at an inclination of $70^\circ$ with an area 
filling factor for the CO emission of $f_{\rm A}=0.5$ (bottom). Curves
are shown for selected LVG models. For the cold component $r_{\rm min}$ has been set based on the radius
of the warm component to 50 and 100\,pc for the
face-on and inclined case respectively. For all other models we used $r_{\rm min}$=1\,pc.
The horizontal lines show the total, observed CO(1--0) line luminosity 
(Riechers \etal \cite{riechers06}, solid line) and the contributions
to $L^\prime_{\rm CO}$ from the cold (dashed line) and the warm (dotted line)
gas components derived in Sec.\,\ref{cotwocomp}. The relevant
intersections between these curves are marked by the arrows for the
four models. Resulting source radii and effective magnifications 
are given in the plot.
 }
\label{rmag}
\end{figure}

We note that the above approach assumes that the caustic of the gravitational lens is roughly 
centered on the AGN and that we indeed trace gas circulating the AGN. This approximation is justified as the
observed line width of 480\,\kms\ is most likely dominated by the rotation of the molecular gas and
the symmetry of the line profile suggests that the velocity profile is not differentially magnified.
The radial dependence of $L^\prime_{\rm CO}$ for this simple geometry is shown 
for selected LVG models in Fig.\,\ref{rmag} (top). The intersection of these
curves with the total observed CO luminosity 
and the relative contribution from the warm and cold gas component 
derived in Sec.\,\ref{cotwocomp} corresponds to the 
true source radius. For example, for an equivalent (magnified) 
source size of $r_0=1150$\,pc (from our 2-component CO and dust
models, for a random mixture of warm and cold gas), the intersection is at $r_{\rm true}=105$\,pc 
with an effective magnification of $m_{\rm eff}=120$ --- the same as the optical/IR magnification. 
We also obtain similar magnifications and source radii if we assume both
components are spatially separated, with the hotter gas closer to the
AGN. With $r_0=595$\,pc for the warm gas, we get $r_{\rm
true}=60$\,pc and $m_{\rm eff}=95$. For the cold gas with
$r_0=995$\,pc and taking $r_{\rm min, cold}=50$\,pc
we get $r_{\rm true}=105$\,pc and $m_{\rm eff}=115$.

\noindent The true source radii, however, will be larger if the CO disk is inclined
and the gas area filling factor is smaller than unity. 
For this more realistic geometry Eq. \ref{lcoface} becomes
\begin{equation} 
\label{r-lco-relation}
L^\prime_{\rm CO}(r) = \,\frac{L^\prime_{\rm CO\,obs}}{\pi\,r_0^2}\,
f_{\rm A}\,\int\limits_{r_{\rm min}}^{r} \phi(r')\,m(r')\,r'\,dr'   \ \ \,
\end{equation} 
where $f_{\rm A}$ is the area filling factor of the CO emitting gas and
$\phi(r')$ denotes the arc length of the intersection between a circle of
constant magnification and the projected inclined disk:
\begin{eqnarray}
\phi(r') & = & 2\pi \ \ \ {\rm for}\ r'\le r_{\rm true}\,{\rm cos}(i)
\nonumber \\ 
\phi(r') & = & 2\pi\,{\rm arcsin}\bigg(\frac{r_{\rm true}}{r'}
\sqrt{\frac{r_{\rm true}^2-r'^2}{r_{\rm true}^2-r_{\rm true}^2\,{\rm
cos}^2(i)}}\bigg)  \nonumber \\
& & \ \ \ {\rm for}\  r_{\rm true}\,{\rm cos}(i) < \ r'\le r_{\rm
true} \nonumber \\
\phi(r') & = & 0 \ \ \  {\rm for}\ r'\ge r_{\rm true}
\end{eqnarray}
\\
The effective magnification is then given by
\begin{equation} 
m_{\rm eff} = \frac{r_0^2}{f_{\rm A}\,{\rm cos}({\rm i})\,(r_{\rm true}^2-r_{\rm min}^2)} \ \ \ .
\end{equation}
As second case we show source radii and magnifications for an inclination of $i =70^\circ$ 
and a gas area filling factor of $f_{\rm A}=0.5$ in Fig.\,\ref{rmag}
(bottom). For the filling factor we use here J1148+5251 at $z=6.4$ 
as a template since it is the only high-z source for which 
multiple CO transitions as well as spatially resolved CO maps exist 
in literature (Bertoldi \etal \cite{bertoldi03}, Walter \etal
\cite{walter03}, \cite{walter04}) which allow to estimate the gas filling factor. 
In this geometry the true source radius for the mixed model
($r_0=1150$\,pc) is $r_{\rm true}=275$\,pc  with an effective
magnification of $m_{\rm eff}=100$. For the spatially separated warm
and cold gas models the corresponding radii and magnifications 
are $r_{\rm true}=135$\,pc with $m_{\rm eff}=105$ (warm gas,
$r_0=575$\,pc, $r_{\rm min}=1$\,pc) and
$r_{\rm true}=325$\,pc with $m_{\rm eff}=60$ (cold gas,
$r_0=995$\,pc, $r_{\rm min}=100$\,pc). 

Hence the true source size, which dominates the observed continuum and line emission, is very 
compact with $r_{\rm true}=100-350$\,pc and has a large effective magnification of $m_{\rm eff}=60-110$ 
for a large range of geometries. A caveat is the unknown structure of the underlying CO
distribution (expressed above as the filling factor of the CO emitting gas and 
the inclination). An arbitrary CO distribution with a low area filling factor could extend well 
beyond the strongly magnified region. Thus the above estimates do not 
rule out that the compact, strongly magnified source is surrounded by additional, unlensed gas at larger radii. 
E.g. a molecular reservoir similar to that seen around M82's starburst disk (Walter \etal \cite{walter02}) or 
the spiral arms surrounding the nucleus of NGC\,1068 (e.g. Schinnerer \etal \cite{schinnerer00}) would 
remain undetected as its contribution to the observed CO line intensities would be negligible. We note, however,
that our analysis excludes an additional gas reservoir of several kpc size (as proposed by Papadopoulos \etal \cite{papado01}) 
as it would have a significant contribution to the low--$J$ CO lines (see also Riechers \etal \cite{riechers06}). 

\begin{figure} \centering
\includegraphics[width=8.5cm]{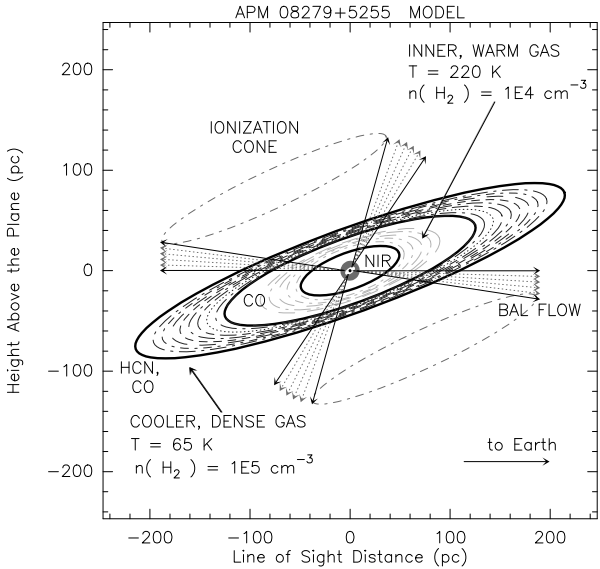}
\caption[Schematic diagram]
{
Schematic diagram of our 2-component model for the CO, HCN, and
mm-FIR dust emission from APM\,08279+5255.  There are two constraints
on the geometry: 1) the large CO and HCN linewidths of 480\,\kms\
imply the molecular rings are being viewed at high inclination.  2) Our
line of sight to the black hole must intersect the BAL outflow cone,
as in the model by Elvis (\cite{elvis00}), so that UV broad absorption lines are
seen against the UV continuum and UV emission lines of the accretion disk.
In the molecular rings, the HCN and high-$J$ CO lines mainly come from the
``cold'', dense component at $>100$\,pc (outer disk), and some of the
mid-$J$ CO emission comes from the ``warm'', lower-density component at
50-100\,pc (inner disk).  The NIR radiation (Egami \etal \cite{egami00}; Soifer \etal \cite{soifer04}) 
comes from the 1500\,K-dust sublimation radius at $\sim 1$\,pc.  
The absorbers responsible for the X-ray BALs are located at radii of $<0.1$\,pc 
from the black hole (Chartas \etal \cite{chartas02}; Hasinger \etal \cite{hasinger02}). 
}
\label{schema} 
\end{figure}

\noindent At a true radius of $\sim 100-250$\,pc, the lens image in the
Egami \etal model is an Einstein ring or filled circle, which may explain the
observed symmetric CO(9--8) distribution.  Egami \etal predict a
diameter $d\approx0.5''$ for this ring --- somewhat smaller than that
derived from the CO(9--8) $u,v$ fits. Thus our high resolution data
also hint to a overall CO size out to $\approx300$\,pc. A possible contradiction to 
this prediction is the morphology of the CO(1--0) distribution observed by Lewis \etal 
(\cite{lewis02a}), which is an incomplete ring. New high-resolution
VLA CO(1--0) observations with better signal-to-noise ratio,
however, show that the CO is very similar to the IR distribution
(Riechers \etal in prep.), supporting the idea that the CO
magnification is higher than the values derived in other studies.
The results of the 2-component modeling for CO, HCN, 
and the dust continuum are summarized in Table\,\ref{model}
and in the schematic diagram in Fig.~\ref{schema}.

\begin{table*}
\caption{Summary of Best-fit Model}
\label{model}
\begin{tabular}{l c c c c l l}
\hline
                        &       &     &``Cold''    &``Warm'' \\
{\bf Parameter} 	&Symbol &Unit &      &      &Remark\\
                        &in text &    &      &\\
\hline
\multicolumn{2}{l}{ {\bf Apparent (Lensed) properties:} }
\\
Lens magnification         &$m$   &---   &$60-110$ &$\sim100$ &model from Egami \etal (\cite{egami00})
\\
Effective magnified radius &$r_0$ &pc &995       &575   
                                                &
\\
CO Luminosity &$m\,L^\prime_{\rm CO(1-0)}$ 
              &\Kkmspc  &$7.4\cdot 10^{10}$  &$3.0\cdot 10^{10}$
\\  
FIR luminosity    &$m\,L_{\rm FIR}$ &\Lsol
                         &$2\cdot 10^{13}$  &$1.7\cdot 10^{14}$	
\\
Apparent dust mass       &$m\,M_d$  &\Msun 
                         &$2.6\cdot 10^9$   &$2\cdot 2\cdot 10^8$   
\\
$M_{\rm gas}$ from CO    &$m\,M$(H$_2$+He)  &\Msun     
                   &$3.7\cdot 10^{11}$       &$2.4\cdot 10^{10}$  
\\
$M_{\rm gas}$ from HCN    &$m\,M$(H$_2$+He)  &\Msun     
                   &$2\cdot 10^{11}$       &--- & contribution from collisional excitation only
\\
$M_{\rm gas}$ from dust    &$m\,M$(H$_2$+He)  &\Msun     
                   &$3.9\cdot 10^{11}$       &$3\cdot 10^10$
\\
 $M_{\rm gas}$ from lvg model    &$m\,M$(H$_2$+He)  &\Msun     
                   &$4\cdot 10^{11}$       &$1.3\cdot 10^{10}$
\\ 
\\
\multicolumn{2}{l}{ {\bf Intrinsic CO properties:} } 
\\
Gas kinetic temperature  &$T_{\rm kin}$    &K   &65   &220  & 
\\
Gas density   &$n({\rm H}_2)$   &cm$^{-3}$    &$1\cdot 10^5$   &$1\cdot 10^4$
\\
CO(1--0) brightness temp. &$T_b$        &K   &$50$   &60     
\\
True CO luminosity &$L^\prime_{\rm CO}$     &\Kkmspc  
                   &$9.2\cdot 10^8$     &$1.3\cdot 10^8$  & CO(1--0); $m_{\rm cold}$=80, $m_{\rm warm}$=100
\\
Gas mass           &$M$(H$_2$+He)  &\Msun     
                   &$5\cdot 10^9$     &$1\cdot 10^8$  & $m_{\rm cold}$=80, $m_{\rm warm}$=100
\\

True CO radius     &$r_{\rm true}$      &pc   
                   &$100-350$         &$65-150$ 
\\
Dynamical mass within $r$     &$rV^2_{\rm rot}/G$  &\Msun   
                   &$1.6\cdot 10^{10}$     &$6.5\cdot 10^9$  
		   &for i=65$^\circ$, $r_{\rm cold}$=250\,pc, $r_{\rm warm}$=100\,pc 
\\
\\
\multicolumn{2}{l}{{\bf Intrinsic Dust quantities:}}
\\
Dust Temperature   &$T_d$   &K   &$65\pm 18$     &$220\pm 30$ 
\\
True FIR Luminosity &$L_{\rm FIR}$      &\Lsol       
                    &$2.5\cdot 10^{11}$   &$1.7\cdot 10^{12}$  & $m_{\rm cold}$=80, $m_{\rm warm}$=100
\\  
True dust mass      &$M_d$              &\Msun 
                    &$3.3\cdot 10^7$    &$2.1\cdot 10^6$   &$m_{\rm cold}$=80, $m_{\rm warm}$=100\\
\hline
\end{tabular}
\end{table*}

\section{Discussion}
\subsection{Heating of the molecular gas}
The properties of the "warm" gas component are best explained by gas heated by the AGN
itself. For an intrinsic luminosity of  $L_{\rm bol}=5\cdot10^{13}\,\lsol$ the expected
distance from the AGN for dust at $\sim200$\,K is $\sim40$\,pc which is consistent within
the uncertainties of the gas geometry with the average distance of the warm component derived above
($r_{\rm min}\approx1$\,pc, $r_{\rm max}\approx60-130$\,pc). This view is also supported by the dust spectrum
which has the typical signature of a hot AGN torus (no PAH features, Soifer \etal \cite{soifer04}).
At distances out to $\sim 350$\,pc the heating of the AGN has dropped to $\sim65$\,K. This temperature
has to be regarded as an upper limit as we here ignore the effect of self--screening in the molecular
toroid. Without more detailed models on the dust heating from the AGN we can not rule out that also the 
"cold" gas component could be substantially heated by the AGN. We can, however, get an independent estimate 
on the heating source for the cold gas using our HCN analysis and the
$L_{\rm IR}$--$L'_{\rm HCN}$ relation observed in local galaxies (Gao \& Solomon \cite{gao04}; 
recalculated to FIR-HCN by Carilli \etal \cite{carilli05}). From the HCN models the HCN(1--0) luminosity
is $L^\prime_{\rm HCN}$ = $5\cdot 10^{10}\,m^{-1}$\,\Kkmspc\ out of which $L^\prime_{\rm HCN}$ = $2\cdot 10^{10}\,m^{-1}\,$\,\Kkmspc\
are due to collisional excitation. The later value corresponds to a FIR luminosity of $1.6\cdot 10^{13}\,m^{-1}\,\lsol$
which is close to the FIR luminosity derived for the cold gas component from the two 
component dust fit ($2\cdot 10^{13}\,m^{-1}\,\lsol$). Therefore the heating of the "cold" gas is probably dominated by
star formation. Taking the magnification of the gas into account ($m_{\rm cold}\approx80$) the implied star 
formation rate in the cold gas component is only $\sim 25\, \msol\,{\rm yr}^{-1}$.
From our decomposition of the dust SED and the above arguments it is therefore clear that 
the FIR luminosity in APM\,08279 is mainly powered by the AGN and
that the contribution of active star formation to the total observed $L_{\rm FIR}$
is only of order $\sim10$\%. \\
The radial temperature profile expected from the AGN heating also implies that it is unlikely that the 
cold and the warm gas are randomly mixed but it supports the interpretation that the cold gas arises
from regions at larger distances from the AGN. As the cold gas dominates the HCN(5--4) line 
luminosity this is also supported by the somewhat narrower linewidth observed in the HCN(5--4) line. Further
support for this picture comes from the analysis of the IR pumping of HCN which shows that the cold
gas is only weakly exposed to the 220\,K IR dust field (IR$_{\rm ff}<1\%$). Such a low filling factor
of the IR field contradicts the view that the cold gas is embedded in the warm gas phase.

\subsection{Estimates of the molecular gas mass.\label{gasmass}}
We now establish estimates of the molecular gas mass using the CO, HCN and
dust emission. For this we adopt the CO(1-0) and HCN(1-0) line
luminosities derived in Sect. \ref{cotwocomp} and \ref{irpump}:
$L^\prime_{\rm CO}$(cold) = $7.4\cdot10^{10}$\,\Kkmspc, 
$L^\prime_{\rm CO}$(warm) = $3.0\cdot10^{10}$\,\Kkmspc\ and 
$L^\prime_{\rm HCN}$ = $2\cdot 10^{10}$\,\Kkmspc.\\
For high-$z$ objects with high FIR luminosities,
a ULIRG conversion factor of 0.8\,\msol(\kkmspc)$^{-1}$ (Downes \&
Solomon \cite{downes98}) is typically adopted to convert the lower-$J$ CO
line luminosities to the total molecular gas mass (see e.g. Solomon \&
Vanden Bout \cite{solomon05}).  The original argument for using this
factor is that much of the CO emission in the rapidly-rotating
circumnuclear disks of ULIRGs comes from a spread-out intercloud
medium, not from self-gravitating clouds (Downes \etal
\cite{downes93}).  As we have seen from our two component gas model, 
even the warm component has a \hh\ density typical for star forming galaxies 
(n(\hh)$\approx 10^4\,{\rm cm}^{-3}$) and about 70\% of the CO(1--0) emission 
comes from extremely dense gas. Unlike typical CO lines from ULIRGS, the global CO emission in APM\,08279 
is therefore not dominated by a diffuse intercloud medium, but by very dense gas, with $n(\hh)=10^5$\,cm$^{-3}$. 
The conversion factor $\alpha$ scales as $n^{0.5}/T_b$,
both for self-gravitating molecular clouds (see the typical ranges of the 
factor $\alpha$ listed by Radford \etal (\cite{radford91}), and for a more distributed 
medium that is not self-gravitating (Downes \etal \cite{downes93}).
For the cold gas component of APM\,08279, the CO brightness
temperatures is similar to those 
in ULIRGS ($T_b\approx\,50$\,K), but the {\it gas density} is about
forty times higher than in ULIRGS. This suggests the conversion
factor should be $\alpha$  $\approx\ (40)^{0.5} \times 0.8$  $\approx\ 6$\,\msol(\kkmspc)$^{-1}$. 
This gives us a first estimate of the molecular gas mass, based on CO alone:
\\
(1) From the CO luminosity of the cold gas component and 
the ``scaled ULIRG factor'' of 5\,\msol(\kkmspc)$^{-1}$ 
we obtain $M_{\rm cold} = 3.7\cdot 10^{11} m^{-1}\,\msol$. Using
the standard ULIRG factor for the warmer gas adds only
$2.4\cdot 10^{10} m^{-1}\,\msol$. gas mass leading to 
$M_{\rm total} = 3.9\cdot 10^{11} m^{-1}\,\msol$.
This gas mass should be compared to estimates from HCN and from the dust:
\\
(2) For the HCN(1--0) line luminosity, Gao
\& Solomon (\cite{gao04}) suggest a conversion factor of
$\sim\,10\,\msol\,$(\kkmspc)$^{-1}$. This yields a value of 
$M_{\rm cold} = 2\cdot 10^{11} m^{-1}\,\msol$. 
\\
(3) From our dust models we derive a dust mass of the cold component of  
$2.6\cdot 10^{9}\,m^{-1}\,\msol$.  If we multiply this dust mass 
by a gas-to-dust mass ratio of 150, we obtain a gas mass of 
$M_{\rm cold}$ =$3.9\cdot 10^{11}\,m^{-1}\,\msol$, 
the same as the estimate from the CO but a factor of two higher than the estimate from HCN. 
With the same gas-to-dust mass ratio the warm dust only adds $M_{\rm warm}$ =$3\cdot 10^{10}
m^{-1}\,\msol$ leading to $M_{\rm total} = 4.2\cdot 10^{11} m^{-1}\,\msol$.
The mass estimate from the dust could be lower, if metals have been 
enriched to a value 2 to 5 times solar, as
suggested by X-ray observations of the iron K-shell absorption edge
in APM~08279 (Hasinger \etal \cite{hasinger02}).\\
If, however, we accept the mass estimate from the optically thin part
of the dust spectrum as a good working estimate,
we can compare this mass directly with the CO and HCN luminosities, to 
``derive'' the conversion factor $\alpha$ for the cold dense gas. This 
directly gives
$\alpha_{\rm CO} = 5.2\,$\msun \,(\Kkmspc)$^{-1}$ and 
$\alpha_{\rm HCN} = 20\,$\msun\,(\Kkmspc)$^{-1}$.  $\alpha_{\rm CO}$ 
agrees well with our value estimated by scaling the ULIRG factor. The 
conversion factor for HCN is a factor of 2 higher than the values 
given in Gao \& Solomon (\cite{gao04}) by in agreement for those
derived in Arp\,220 and NGC\,6240 (Greve \etal \cite{greve06}).
\\
(4) We can also estimate the gas mass from the LVG models rather than 
using a conversion factor (see Eq.\ref{lvgmasseq}):. Assuming that the low [CO]/\gradv\ we use in
our models is mainly due to a higher turbulence and not due to a lower
CO abundance, \gradv\ is $\sim 5\,\kms\,{\rm pc}^{-1}$. For the turbulence line
width we adopt here  $\Delta\,V_{\rm turb}=100\,\kms$ based on estimates for
SMM\,J02399-0136 ($z=2.8$) where the ratio of the local velocity 
dispersion to rotation velocity has been determined to be $\sim\,0.2$ (Genzel \etal \cite{genzel03}).
Similar turbulence motions have been derived for Arp\,220 (Downes \& Solomon \cite{downes98}).
With this number the gas mass from the LVG model becomes 
$M_{\rm cold}=4\cdot 10^{11}\,m^{-1}\,\msol$ and  $M_{\rm warm}=1.3\cdot 10^{10}\,m^{-1}\,\msol$  in agreement with the
estimates above. With these numbers the equivalent path length
$\Delta\,V/\gradv$\ through the molecular disk is $\sim20$\,pc.
\\
(5) Another check on the minimum gas mass is to simply calculate 
the CO optically thin limit.  Following Solomon \etal (\cite{solomon97}), 
this is 
\begin{equation}
M_{\rm thin}({\rm H}_2+{\rm He}) = 9.4\cdot 10^{-3} L^\prime_{\rm CO} T_{\rm ex} \ \ ,
\end{equation}
where $L^\prime_{\rm CO}$ is the CO(1--0) line luminosity  in \Kkmspc ,
$T_{\rm ex}$ is the excitation temperature in K, $M_{\rm thin}$
is a lower limit to the gas mass in \Msun , 
and the numerical coefficient is for an assumed (Milky Way) CO abundance
of [CO]/[H$_2$] of 10$^{-4}$.
Thus for the cold molecular gas in APM\,08279 with a CO abundance of
$5\cdot10^{-5}$ the lower limit to the gas mass is $M_{\rm thin}=5.2\cdot 10^{10}\,m^{-1}\,\msol$ .\\
In summary, the gas mass estimates from the CO, HCN, the optically-thin
millimeter dust continuum, and the CO radiative-transfer modeling 
are all in reasonable agreement and lead to a total gas mass of 
$M_{\rm total}=4\cdot 10^{11}\,m^{-1}\,\msol$. The gas mass is dominated
by the cold dense HCN emitting gas. Taking the magnification of $m_{\rm cold}\approx80$ into account
the intrinsic gas mass of APM\,08279 is $\sim5\cdot10^{9}\,\msol$. 
The implied CO luminosity-to-gas mass conversion factor is about 5 to 6\,\msol\,(\kkmspc)$^{-1}$. 

\subsection{Mass budget in the central region of APM\,08279}
\begin{figure}[h] 
\centering
\includegraphics[width=8.5cm]{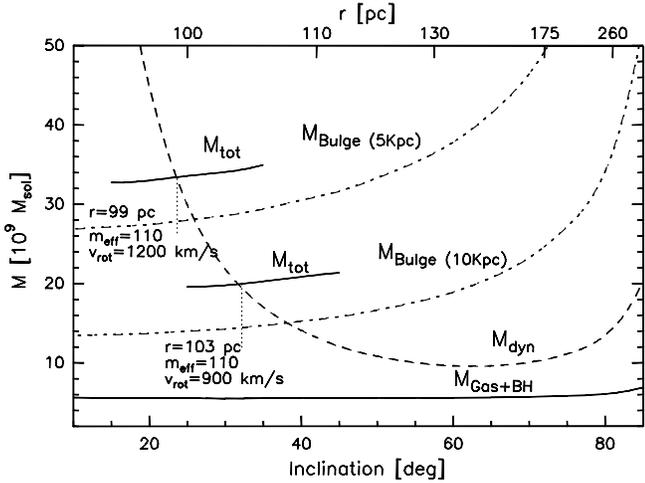}
\caption[Mass Contributions]
{Black hole, molecular, stellar and dynamical mass in the
central region of APM\,08279 as a function of assumed inclination of
the molecular disk. Radii as a function of the inclination have been
calculated from Eq. \ref{r-lco-relation} for the cold, dense gas
($r_0=995$\,pc, $r_{\rm min}=1$\,pc) using a filling factor of unity. The stellar mass
has been calculated using a scalelength of the stellar bulge of 5 and
10\,kpc and assuming the $M_{\rm BH}-\sigma\!_*$ relation were
to hold. Variations of the  stellar bulge contribution with
inclination are only due to changes of the size of the region 
considered. The corresponding radii are given at the top axis of the plot.
The inclination for which the total mass (gas\,+\,stellar\,+\,BH)
matches the dynamical mass are given for both bulge geometries
together with the corresponding numbers for the true CO radius,
effective magnification of the molecular gas, and the 
CO rotation velocity.
}
\label{mass-geo}
\end{figure}

Having established estimates for the intrinsic size of the CO emitting
region (Sect. \ref{magsection}) and for the molecular gas mass (Sect.
\ref{gasmass}) we now discuss the mass budget in the central
region of APM\,08279 to test if it falls on the $M_{\rm BH}-\sigma\!_*$ relation
found for local galaxies (Gebhardt \etal
\cite{gebhardt00}, Ferrarese \& Merritt \cite{ferrarese00}). Shields \etal (\cite{shields06})
have recently addressed this question for a sample of high-$z$ QSOs,
including APM\,08279. For APM\,08279 those authors assume a much lower 
magnification (7) than we do ($\sim 100$), and consequently assume the 
CO is spread out over a 2-kpc diameter.  In the present paper, 
we concentrate the discussion toward the central few hundred-pc
region:\\
The bolometric luminosity of the quasar, after correcting for the
factor-of-100 amplification by gravitational lensing, is still an
enormous $7\cdot 10^{13}$\,\Lsol. From the argument originally due 
to Zel'dovich \& Novikov (1964), the lower limit to the black hole 
mass, if electron scattering produces the main opacity, is
\begin{equation}
M_{\rm BH} \geq  3.0\cdot 10^{-5} L_{\rm bol} \ \ \ ,
\end{equation} 
with $M_{\rm BH}$ and $L_{\rm bol}$ in solar units. This means the mass of the
black hole at the center of APM\,08279 likely exceeds $2\cdot10^9$\,\Msun.
If the local $M_{\rm BH}-\sigma\!_*$ relation were to hold, then the
stellar bulge would have a mass of $M_{\rm bulge}=1.4\cdot10^{12}$\,\Msun.
To estimate the stellar mass in the central few hundred pc, we assume 
a central density profile for the bulge of $\rho\sim r^{-2}$ (Jaffe
\cite{jaffe83}, Tremaine \etal \cite{tremaine94}) and a bulge scale
length of 5 and 10\,kpc. Since the inclination of the molecular disk
is largely unconstrained, we treat it in the following as a free
parameter. Using Eq. \ref{r-lco-relation}, we have first calculated the 
CO source size as function of the inclination. We use here as an
example only the dense cold gas component from the two component
model, since the warm gas does not contribute significantly to the
mass. We have further used a gas filling factor of unity and $r_{\rm
min}=1$\,pc, since then the CO source is most compact, which
maximizes its gravitational impact in the 
central region. From the CO source size we have then calculated the 
dynamical mass and the stellar mass contribution. The resulting mass
contributions as a function of the inclination are shown in
Fig.\,\ref{mass-geo}. From this figure it is obvious that, independent
of the inclination, the sum of the black hole and the gas mass is
always lower than the dynamical mass. Given that we show a model
with one of the highest central gas mass concentration, this holds for all other
configurations of the two component gas model as well. The figure also
shows that the stellar mass exceeds the dynamical mass for
inclinations larger than $\sim40^\circ$. Note that the inclination for
which the stellar and the dynamical mass intersect is independent of the
assumed gas model since both masses scale linear with the underlying
source radius (which is determined by the gas model). Taking the 
gas and black hole mass into account, the limits for the inclination
of the CO disk become $<25^\circ$ and $<35^\circ$ for a bulge
scale length of 5 and 10\,kpc respectively. Thus, if the
local $M_{\rm BH}-\sigma\!_*$  relation were to hold, the implied
intrinsic rotation velocity of the CO disk would be $v_{\rm rot} >
900\kms$, and $v_{\rm rot} > 1200\kms$. Given that the observed
CO linewidth of 480\,\kms\ is already one of the highest linewidths
observed in local and high-z QSOs (see e.g. Solomon \& Vanden Bout
\cite{solomon05} and references therein), it is more likely that we see
the CO disk in APM\,08279 at high inclination, which implies that
 $M_{\rm BH}-\sigma\!_*$  does not hold for APM\,08279. 

\noindent The conclusion
is that for this QSO at $z=3.9$, the super-massive black hole is 
already in place, but the assembly of the stellar bulge is still in 
progress. Similar conclusions have been gained by Walter \etal
(\cite{walter04}) for J1148+5251 at $z=6.4$ and by Shields \etal (\cite{shields06})
on a larger sample of high-$z$ QSOs.A caveat 
arising from the special lens configuration in APM\,08279 is, that
the emission lines only trace a very compact region surrounding the AGN which 
implies that this conclusion only holds if the central density profile of the stellar 
bulge follows a $\rho\sim r^{-2}$ profile to the central $\sim100$\,pc.

\subsection{Comparison with other high-$z$ QSOs}
\begin{figure}[h] 
\centering
\includegraphics[width=8.5cm]{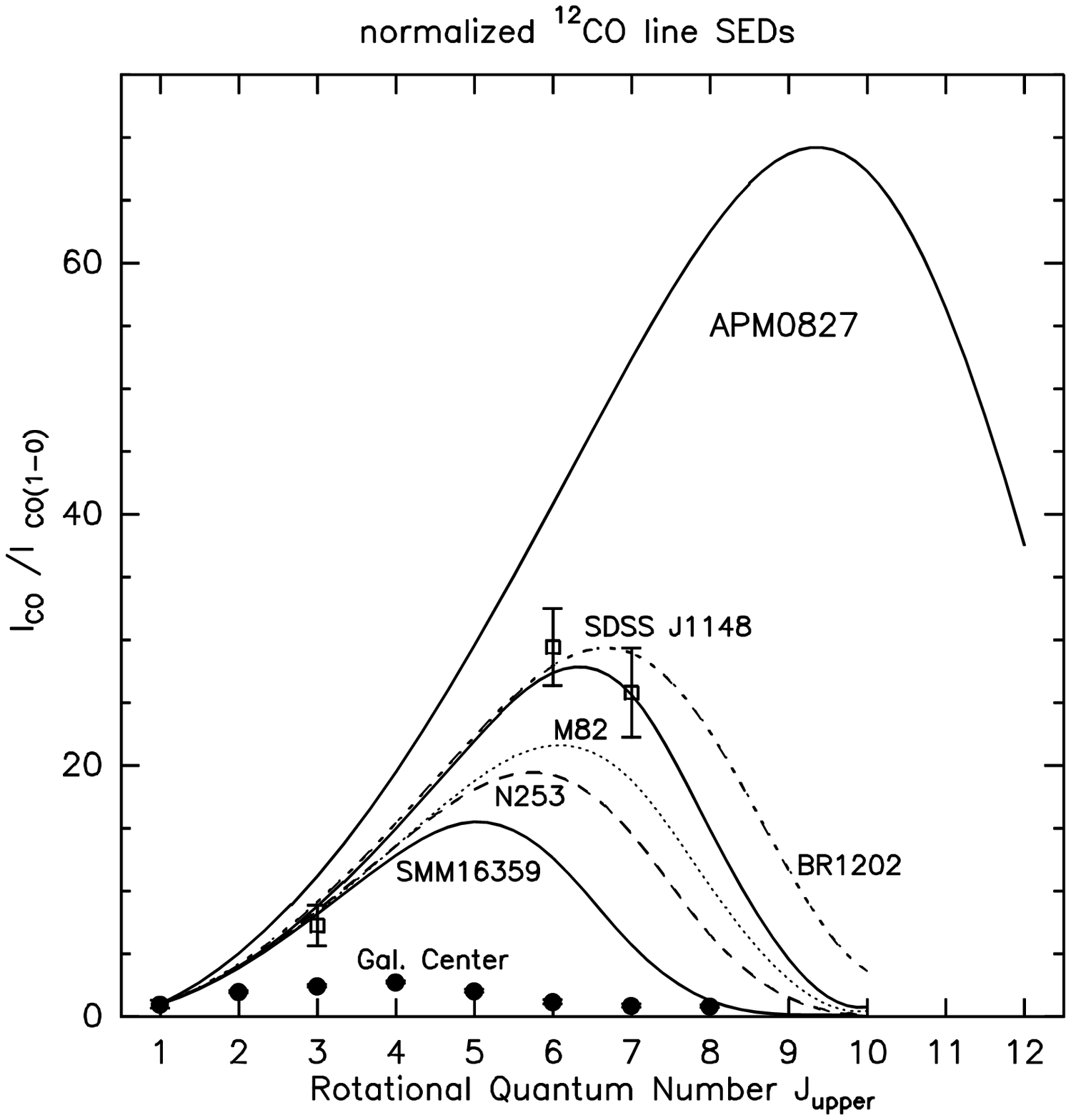}
\caption[CO SED]
{
Comparison of the CO line SEDs of selected local and high-z galaxies.
The SEDs are shown for APM\,08279 (this paper, Fig.\ref{cosed}),  
BR1202-0725 ($z=4.7$, Carilli \etal \cite{carilli02}, Riechers \etal \cite{riechers06}), J1148+5251
($z=6.4$, open squares, Bertoldi \etal \cite{bertoldi03}, Walter \etal, \cite{walter03})
the high-excitation component in the center of M82 (Wei\ss\ \etal \cite{weiss05b}),
NGC\,253 center (G\"usten \etal \cite{guesten06})
SMM\,16359 ($z=2.5$, Wei\ss\ \etal \cite{weiss05a}) and 
the Galactic Center (solid circles, Fixsen \etal \cite{fixsen99}). The CO line SEDs
are normalized by their CO(1--0) flux density.
}
\label{cosedall} 
\end{figure}

Our observations show that the CO line SED of APM\,08279 is
strikingly different from those of other high-$z$ QSOs and
nearby starburst galaxies (Fig.\,\ref{cosedall}). The turnover of the CO SED in high-$z$ QSOs
typically occurs at the CO(6--5) or CO(7--6) transition (e.g. BR1202-0725,
J1148+5251). This also holds  for the F10214 the Cloverleaf and other
high-z QSOs (Wei\ss\ \etal in prep).
In contrast to these sources, the CO line SED turnover in APM\,08279
occurs at the CO(10--9) line!  This is mainly due to the high \hh\ density in excess of
$10^5$\,cm$^{-3}$ which is more than ten times higher
than those in other dusty QSOs. This probably also explains the high luminosity
of the HCN line in APM\,08279, relative to that in other quasars. The
presence of a high-temperature (200\,K) CO and dust component also
distinguishes APM\,08279 from other high-$z$ quasars detected in CO so
far.  From the lens modeling, both the cold component, and
the warm molecular gas component in APM\,08279 appear to be
closer to the nucleus than in the other quasars which makes it likely
that the 200\,K component is heated directly by the AGN.  As noted 
by Blain (\cite{blain99}), part of these differences relative to other quasars may 
be due to the configuration of the gravitational lens, which 
gives us a high-magnification zoom right into the central 100--300 pc radius
of APM~08279.  If such an inner circumnuclear region of high-density 
molecular gas exists in all the high-$z$ quasars detected in 
molecular lines so far, then it might be relatively inconspicuous, 
because it is not highly magnified, and it has a low filling factor 
relative to the greater circumnuclear region out to a radius of 1\,kpc.
Conversely, in APM\,08279, we may be missing the gas out to 1\,kpc,
because the central region outshines the more extended gas due to its high flux 
amplification.  Although the Egami \etal (\cite{egami00})
lensing model indicates the central region may be free of 
differential magnification, we still may not even see the outer 1\,kpc-radius
region, simply because its magnification has fallen by nearly a factor of 100.

\section{Conclusions}
We observed APM\,08279 as part of our ongoing CO line SED
study of high-$z$ sources at the 30\,m telescope. 
Besides the previously observed CO(4--3) and (9--8) lines, new detections 
included the CO(6--5), (10--9) and the (11--10) lines.
We also present improved observations of the CO(4--3), CO(9--8), 
and HCN(5--4) lines and the 1.4 and 3mm  dust continuum with the IRAM Interferometer. 
The simple fact that such high CO lines are detected 
shows that the gas
properties of APM\,08279 differ greatly from those seen in other high-$z$
QSOs and local starbursts. \\
The CO and dust size measurement, combined with the lensing model of 
Egami \etal (\cite{egami00}) shows that the source region is very compact 
(radius $\sim$ 60--300\,pc) and that its magnification is similar to 
the optical/IR magnification ($\sim 60-110$). The high CO excitation and 
the dust continuum emission are best modeled with a 2-component gas
model. The warm ($\sim 220$\,K) gas component is most likely directly heated 
by the AGN and arises from radii between 60--150\,pc. This component
dominates the FIR luminosity ($\sim 90\%$). The cooler gas ($\sim 65$\,K),
which carries $\sim 90\%$ of the total gas mass of 
$M_{\rm gas}\approx5\cdot 10^9$\,\msol\ arises from somewhat
larger radii. Its main characteristic is a high
H$_2$ density of $\sim 1\cdot 10^5$\, cm$^{-3}$ -- about 10 times
higher than the gas in other high-z QSOs and local starburst
galaxies. The high gas density implies that the standard ULIRG conversion factor,
usually applied to high-z galaxies, does not provide a good estimate
of the molecular gas in APM\,08279. Our study suggests that conversion
factor for CO, in this special case, is 
$\alpha_{\rm CO} \approx 5\,$\msun \,(\Kkmspc)$^{-1}$.\\
Although the gas density for the cold gas is typical for the HCN emitting volume in 
local ULIRGs it is not high enough to explain the observed HCN(5--4) luminosity
in APM\,08279. From LVG models including IR pumping via the 
14\microns\ bending mode of HCN we conclude that the hot dust heated by the AGN
efficiently boosts the HCN excitation and that this excitation channel is more important than the
collisional excitation for this particular source. The star formation rate associated 
with the dense gas component is only $\sim 25\, \msol\,{\rm yr}^{-1}$.\\
An estimate of the mass budget in the central region of APM\,08279
suggests that the black hole and gas mass does not leave sufficient room
for a stellar mass contribution following the local $M_{\rm
BH}-\sigma\!_*$ relation for reasonable inclinations of the molecular
disk.
Hence we conclude that the super-massive black hole in APM\,08279 is 
already in place, before the buildup of the stellar bulge is complete.

\begin{acknowledgements}
We thank the IRAM receiver engineers D.~John and S.~Navarro 
for their great help in optimizing receiver tunings as well as
the 30\,m telescope operators and the Plateau de Bure Interferometer 
operators for their all-around assistance with the observing in general. 
IRAM is supported by  INSU/CNRS (France), MPG (Germany) and IGN (Spain). 
\end{acknowledgements}


\end{document}